\documentclass{jfm}

\usepackage[english]{babel}
\usepackage[utf8x]{inputenc}
\usepackage[T1]{fontenc}

\usepackage{amsmath}
\usepackage{graphicx}
\usepackage{subcaption}

\usepackage{float}
\usepackage{graphicx}
\usepackage{epstopdf, epsfig}
\usepackage{flushend}
\usepackage[dvipsnames]{xcolor}
\usepackage{amsmath}
\usepackage{latexsym}
\usepackage{caption}
\usepackage{color,soul}
\usepackage{physics}
\usepackage[normalem]{ulem}

 \usepackage{physics}

\graphicspath{{figures/}}
\usepackage[colorlinks=true, allcolors=blue]{hyperref}
\usepackage{xcolor}
\usepackage{pgfplots}
\usepackage{bm}
\usepackage{todonotes}
\usepackage{comment}

\newcommand{\ethan}[1]{\textcolor{black}{#1}}

\newcommand{\Revthree}[1]{\textcolor{black}{#1}}

\usepackage{algorithm,amsmath,algpseudocode,algcompatible,setspace}

\newcommand\solidrule[1][0.5cm]{\rule[0.5ex]{#1}{1pt}}
\newcommand\dashedrule{\mbox{\solidrule[1mm]\hspace{1mm}\solidrule[1mm]\hspace{1mm}\solidrule[1mm]}}
\newcommand\dottedrule{\mbox{\solidrule[0.5mm]\hspace{0.5mm}\solidrule[0.5mm]\hspace{0.5mm}\solidrule[0.5mm]\hspace{0.5mm}\solidrule[0.5mm]\hspace{0.5mm}\solidrule[0.5mm]}}

\title{Optimal eddy viscosity for resolvent-based models of coherent structures in turbulent jets}

\author{Ethan Pickering\aff{1}, Georgios Rigas\aff{1}, Oliver T. Schmidt\aff{2}, Denis Sipp\aff{3}, Tim Colonius\aff{1}}

\affiliation{
\aff{1} Division of Engineering and Applied Science, California Institute of Technology,
Pasadena, CA 91125, USA
\aff{2} Mechanical and Aerospace Engineering, University of California, San Diego, La Jolla, CA 92093, USA
\aff{3} ONERA - The French Aerospace Lab, 92190 Meudon, France
}

\begin{document}
    \maketitle
    \begin{abstract}
Response modes computed via linear resolvent analysis of a turbulent mean-flow field have been shown to qualitatively capture characteristics of the observed turbulent coherent structures in both wall-bounded and free shear flows.  \Revthree{To make such resolvent models predictive, the nonlinear forcing term must be closed. Strategies to do so include imposing self-consistent sets of triadic interactions, proposing various source models, or through turbulence modelling.} For the latter, several investigators have proposed using the mean-field eddy viscosity acting linearly on the fluctuation field.  In this study, a data-driven approach is taken to quantitatively improve linear resolvent models by deducing an optimal eddy-viscosity field that maximizes the projection of the dominant resolvent mode to the energy-optimal coherent structure educed using spectral proper orthogonal decomposition (SPOD) of data from high-fidelity simulations.  We use large-eddy simulation databases for round isothermal jets at subsonic, transonic, and supersonic conditions and show that the optimal eddy viscosity substantially improves the alignment between resolvent and SPOD modes, reaching over 90\% alignment at those frequencies where the jet exhibits a low-rank response.  We then consider a fixed model for the eddy viscosity and show that with the calibration of a single constant, the results are generally close to the optimal one.  In particular, the use of a standard Reynolds-Averaged-Navier-Stokes (RANS) eddy-viscosity resolvent model, with a single coefficient, provides substantial agreement between SPOD and resolvent modes for three turbulent jets and across the most energetic wavenumbers and frequencies.

\end{abstract}

\section{Introduction} \label{sec:Intro}

Resolvent analysis (also known as input/output analysis) determines a volumetric distribution of forcing in the frequency domain that gives rise, when acting in a time-invariant flow, to the most amplified linear response, typically measured in terms of its total kinetic energy. It is an important tool in stability and transition analysis \citep{trefethen1993hydrodynamic,farrell1993optimal,schmid2002stability,jovanovic2005componentwise}, and has more recently been proposed as a reduced-order model of coherent structures in fully-developed turbulence \citep{mckeon2010critical,hwang2010linear}. In the latter context, resolvent analysis can be derived by partitioning the Navier–Stokes equations into terms that are linear and nonlinear with respect to perturbations. Such a rearrangement of the equations is exact, and the equations may be explored without recourse to any further modeling. With varying degrees of formality, similar approaches were proposed in the past \citep{malkus1956outline,michalke1971instability,crighton1976stability,butler1992three}, but increases in computer power that speed up the singular value decomposition (SVD) of the linear operator using direct LU decomposition (multi-frontal algorithms for sparse systems) have allowed a detailed characterization of the resolvent spectrum in several turbulent, canonical wall-bounded \citep{hwang2010amplification,hwang2010linear,mckeon2010critical,sharma2013coherent,moarref2013model} and free shear flows \citep{jeun2016input,schmidt2018spectral}.

At those frequencies where the dominant singular value is significantly larger than the subdominant ones (which we refer to as low-rank behavior), the dominant modes are qualitatively similar to coherent modes extracted from data \citep{schmidt2018spectral}. However, when the response is not low rank, a non-trivial structure of the nonlinear forcing terms may lead to discrepancies between resolvent and observed modes.  Thus, it is necessary to model the nonlinear forcing to attain resolvent analyses that are quantitatively predictive. Previous studies have considered several approaches for modeling the nonlinear forcing in linear analyses.  These include empirical models  \citep{bechara1994stochastic,tam1999jet,cavalieri2011jittering,cavalieri2014coherence,towne2017statistical}, estimation given partial statistics of the response \citep{zare2017colour,towne2020resolvent,martini2020resolvent}, and/or the use of a turbulent, or eddy, viscosity.  An eddy viscosity may be motivated by concepts underlying the triple decomposition \citep{reynolds1967stability,reynolds1972mechanics}, which identifies the Reynolds stresses as acting on the coherent fluctuations (from both the coherent and incoherent fluctuations), even though the phase average used to define the coherent part of the turbulent-viscosity field is ambiguous in unforced turbulent flows. Many studies have applied eddy-viscosity models in the wall-bounded turbulence literature \citep{del2006linear,cossu2009optimal,pujals2009note,hwang2010amplification,hwang2010linear,hwang2016mesolayer,vadarevu2019coherent,hwang2020attached} either through implementation of the \cite{cess1958} model or by estimating the eddy-viscosity field via the Reynolds stresses and the mean shear rate of strain. Similarly, global stability analyses have applied eddy-viscosity models to identify and/or control forced or self-sustained resonances in transitional and turbulent flows \citep{crouch2007predicting,meliga2012sensitivity,mettot2014quasi,sartor2014stability,semeraro2016stochastic,tammisola2016coherent,rukes2016assessment,oberleithner2014impact}. These studies implemented eddy viscosity on an {\it ad hoc} basis, citing improved qualitative agreement or improved integrated energy-densities.

In a more quantitative sense, eddy-viscosity enhanced linear models have also proven useful for assimilating known data to reconstruct observed energy spectra and mean-flow quantities. \cite{moarref2012model} showed that a data-driven, white-in-time forcing could reproduce the DNS-based turbulent energy spectrum and, similarly, \cite{illingworth2018estimating} could match DNS energy spectra using time-resolved velocity measurements. More recently, \cite{towne2020resolvent} showed that incorporating an eddy-viscosity model led to accurate estimates of space-time statistics using partially known data from DNS.  Finally, \cite{pickering2020resolvent} used an eddy-viscosity enhanced resolvent model to reconstruct the large-eddy simulation (LES) acoustic field of transonic and supersonic turbulent jets at a significantly lower rank when compared to their non-eddy-viscosity enhanced computations. Other approaches have implemented eddy-viscosity fields to develop self-consistent models, such as \cite{yim2019self} or \cite{hwang2020attached}, where the former study coupled a harmonically forced, quasi-linear resolvent analysis with RANS equations, citing eddy viscosity as a necessary link between the coherent and incoherent perturbation dynamics.

Although the utility of eddy-viscosity enhanced linear models for turbulent modeling and control has become increasingly apparent, a quantitative assessment of their effect on turbulent structures is lacking, even more, it is unclear which statistics turbulence models should seek to predict.  One appealing target are modes educed by spectral proper orthogonal decomposition (SPOD), as these modes optimally reconstruct the turbulent kinetic energy and represent space-time coherent structures \citep{towne2018spectral}. In fact, the SPOD has a theoretical connection with resolvent analysis. \cite{towne2018spectral} showed that if the resolvent forcing modes, at a given frequency and wavenumber, are mutually uncorrelated, then the resolvent response modes are identical to the SPOD modes. Likewise, discrepancies between the SPOD and resolvent modes imply correlated forcing modes.

\cite{morra2019relevance} applied a similar line of thinking by including an eddy viscosity in their resolvent analysis of turbulent channel flow, showing that the resulting resolvent modes were in greater agreement with the SPOD modes educed from high-fidelity simulation data than resolvent analysis using only molecular viscosity. We extend this approach to turbulent jets, but consider a more general framework. The central question we ask is: how well can the inclusion of an eddy-viscosity model in the resolvent operator approximate the correlations of the forcing cross spectral density tensor? In this approach, an ideal model would render any remaining forcing as uncorrelated, meaning that the resolvent and SPOD modes coincide. We therefore define a data-informed variational problem that seeks an optimal eddy-viscosity field that maximizes the projection of the first SPOD mode on the first resolvent mode. We then show that we can achieve nearly optimal projections using standard eddy-viscosity models, including one directly inferred from a corresponding Reynolds-Average Navier-Stokes (RANS) simulation. 

The work presented here is also relevant to a broader debate taking place regarding the interpretation of resolvent analysis.  Since we can define the resolvent operator from the full nonlinear equations without introducing approximations or closures, it is attractive to proceed without introducing {\it ad hoc} models such as eddy viscosity, since we can still consider the framework {\it exact}. With a minor caveat (i.e. while exact, the resolvent decomposition is not necessarily unique as it can depend on the choice of dependent variables used to express the governing equations \citep{karban2020ambiguity}), this implies that the forcing terms are {\it physically} interpretable (i.e. measurable) quantities.  This perspective is, in our opinion, valuable, and may be pursued alongside efforts (such as the present work) aimed at empirically modeling the forcing. However, there is a subtlety that confounds the separation between ``exact’’ and ``modeled’’ resolvent analyses: namely, it may not be possible to compute, with meaningful accuracy, the exact resolvent modes in high Reynolds number flows, particularly when the mean flow is two- or three-dimensional.  The fine-scale structure of the modes can require resolutions similar to DNS, and inversion of the resulting linear systems for singular value decomposition can be prohibitive.  A survey of resolvent analyses conducted to date on multidimensional base flows show that a variety of {\it regularizations} of the resolvent operator have been used to reduce the computational burden. These include the use of eddy-viscosity models (as discussed at length above), fourth-order numerical filters \citep{jeun2016input}, effective Reynolds numbers \citep{schmidt2018spectral}, and linear damping \citep{yeh2019resolvent}.

From a more general perspective, the present work also has a connection to the building of data-augmented turbulence models \citep{duraisamy2019turbulence}. Here, we specifically target the modeling of unsteady features \citep{wang2018pof,maulik2019jfm} and the optimal eddy-viscosity fields found, at each frequency-wavenumber pair, which are analogous to field-inversion steps (also based on variational data-assimilation methods, \citet{foures2014data,parish2016}) that assist machine learning techniques in generating eddy-viscosity models from mean-flow quantities.

We organize the paper as follows. In \S~\ref{sec:methods} we outline the governing equations, resolvent analysis, and SPOD. In  \S~\ref{sec:models} we discuss the optimization framework developed to align SPOD and resolvent modes, and the specific eddy-viscosity models examined. \S~\ref{sec:results} provides the resulting resolvent mode shapes found via the four eddy-viscosity models and \S~\ref{sec:Opt_params} analyzes the associated optimal eddy-viscosity fields. In \S~\ref{sec:subdominant} we show a favorable impact of the eddy-viscosity models on the subdominant resolvent modes and then conclude the analysis in \S~\ref{sec:complete} by assessing the sensitivity of the RANS eddy-viscosity model. In this final section, we ultimately find a frequency independent RANS eddy-viscosity field that performs well for three turbulent jets (i.e. subsonic, transonic, and supersonic) and their most energetic frequencies ($St \in[0.05,1]$) and azimuthal wavenumbers ($m \in \mathbb{N} \subset [0,5]$).

\section{Methods} \label{sec:methods}

The LES database, resolvent analysis, and SPOD were described in \cite{schmidt2018spectral} and \cite{towne2018spectral}. For brevity, we only recall the main details here.

\subsection{Large Eddy Simulation database}

\begin{table}
    \centering
\begin{tabular}{c c c c c c c c}
 case & $M_j$ & $Re_j$ & $\frac{p_0}{p_\infty}$ & $\frac{T_0}{T_\infty}$ & $n_{\text{cells}}$ & $\Delta t a_\infty/ D$ & $ \Delta St$\\  
 subsonic & $0.4$ & $4.5 \times 10^5$ & 1.117 & 1.03 & $15.9 \times 10^6$ & 0.2 & $0.049$ \\ 
 transonic & $0.9$ & $1.01 \times 10^6$ & 1.7 & 1.15 &$15.9 \times 10^6$ &0.2 & $0.022$ \\  
 supersonic & $1.5$ & $1.76 \times 10^6$ & 3.67 & 1.45 &$31 \times 10^6 $ &0.1 & $0.026$ \\
\end{tabular}
\caption{Parameters, sampling rate, and frequency resolution for the LES.}
    \label{tab:LES}
\end{table}

The flow solver Charles was used to compute the LES databases, including subsonic (Mach 0.4), transonic (Mach 0.9), and supersonic (Mach 1.5) cases; \cite{bres2017unstructured} contains the details on the numerical method, meshing, and subgrid-models.  Experiments conducted at PPRIME Institute, Poitiers, France were used to validate the Mach 0.4 and 0.9 jets \citep{bres2018importance}. Table \ref{tab:LES} provides a summary of parameters for the three jets considered. Parameters include the Reynolds number based on diameter $Re_j = \rho_j U_j D / \mu_j$ (where subscript $j$ specifies the value at the centerline of the jet nozzle exit, $\rho$ is density, $\mu$ is viscosity) and the Mach number, $M_j = U_j/a_j$, where $a_j$ is the speed of sound. The simulated $M_j = 0.4$ jet corresponds to the experiments in \cite{cavalieri2013wavepackets,jaunet2017two,Nogueira2019Streaks} with the same nozzle geometry and similar boundary-layer properties at the nozzle exit. Throughout the manuscript, reported results are non-dimensionalized by the mean jet velocity $U_j$, jet diameter $D$, and dynamic pressure $\rho_j U_j^2$. We report frequencies in Strouhal number, $St = f D / U_j$, where $f$ is the frequency. 

Each database comprises 10,000 snapshots separated by $\Delta t a_\infty/ D$, where $a_\infty$ is the ambient speed of sound, and is interpolated onto a structured cylindrical grid $x,r,\theta \in [0,30] \times [0,6] \times [0, 2\pi]$, where $x$, $r$, $\theta$ are streamwise, radial, and azimuthal coordinates, respectively. Variables are reported by the vector
\begin{align}
\bm{q} = [\rho, u_x, u_r, u_\theta, T]^T,    
\end{align}
where $u_x$, $u_r$, $u_\theta$ are the three velocity components, and a standard Reynolds decomposition separates the vector into mean, $\bar{\bm{q}}$, and fluctuating, $\bm{q}'$, components
\begin{align}
   \bm{q}(x,r,\theta,t) = \bar{\bm{q}}(x,r) + \bm{q}'(x,r,\theta,t).
   \label{eqn:Reynolds}
\end{align}

\subsection{Resolvent analysis}

We start with the nonlinear flow equations of the form
\begin{equation}
    \frac{\partial \bm{q}}{\partial t} = \bm{F}(\bm{q}),
    \label{eqn:NavierStokesCompact}
\end{equation}
where $\bm{F}$ is the time-independent compressible Navier-Stokes operator  (plus continuity and energy). Substituting equation \eqref{eqn:Reynolds} for $\bm{q}$ and separating terms linear in state perturbations, $\bm{q}’$, to the left-hand side gives
\begin{equation}
    \frac{\partial \bm{q}’}{\partial t} - \bm{A}(\bar{\bm{q}})\bm{q}’ = \bm{f}(\bar{\bm{q}},\bm{q}’),
    \label{eqn:LNS}
\end{equation}
where
\begin{equation}
      \bm{A}(\bar{\bm{q}}) = \frac{\partial \bm{F}}{\partial \bm{q}}(\bar{\bm{q}})
\end{equation}
is the linearized flow operator (provided in Appendix \ref{App:LNS})  and $\bm{f}$ contains the nonlinear terms and any additional external inputs (e.g. environmental noise or perturbations at the boundary).

For the round, statistically-stationary turbulent jets we consider, equation \eqref{eqn:LNS} is Fourier transformed both temporally and azimuthally to the compact expression
\begin{equation}
(i\omega\textbf{I} - \textbf{A}_m) \bm{q}_{m, \omega} = \bm{f}_{m, \omega},
\label{eqn:LNS_eq}
\end{equation}
where $\omega = 2 \pi St$ is the frequency and $m$ represents the azimuthal wavenumber. We can then rewrite equation \eqref{eqn:LNS_eq} by defining the resolvent operator, $\bm{R}_{\omega,m} = (i\omega \bm{I} - \bm{A}_m)^{-1}$,
\begin{equation}
    \bm{q}_{m, \omega} = \bm{R}_{m, \omega} \bm{f}_{m, \omega},
    \label{eqn:resolvent}
\end{equation}
and introduce the compressible energy norm  \citep{chu1965energy} via the matrix $\bm{W}$,
\begin{equation}
    \langle \bm{q}_1, \bm{q}_2 \rangle_E = \int \int \int \bm{q}_1^* \text{diag} \bigg( \frac{\bar{T}}{\gamma \bar{\rho} M^2}, \bar{\rho}, \bar{\rho}, \bar{\rho}, \frac{\bar{\rho}}{\gamma (\gamma - 1) \bar{T} M^2} \bigg) \bm{q}_2 r \dd r \dd x \dd \theta = \bm{q}_1^* \bm{W} \bm{q}_2,
    \label{eqn:energy_norm}
\end{equation}
 to the forcing and response, where $\bm{W} = \bm{W}_f = \bm{W}_q$. The resolvent modes under this norm are then found by taking the singular value decomposition of the weighted resolvent operator,
 \begin{equation}
     \tilde{\bm{R}}_{m, \omega} = \bm{W}_q^{1/2} \bm{R}_{m, \omega}\bm{W}_f^{-1/2}  = \tilde{\bm{U}}_{m,\omega} \bm{\Sigma}_{m, \omega} \tilde{\bm{V}}_{m, \omega}^*,
 \end{equation}
 where the diagonal matrix $\bm{\Sigma}_{m, \omega}$ contains the ranked gains and the columns of $\bm{U}_{m,\omega} = \bm{W}_q^{-1/2} \tilde{\bm{U}}_{m,\omega}$ and ${\bm{V}}_{m,\omega} = \bm{W}_f^{-1/2} \tilde{\bm{V}}_{m,\omega}$ contain the response and forcing modes, respectively. These modes are orthonormal in the energy norm, equation \eqref{eqn:energy_norm},
\begin{equation}
    \bm{U}_{m,\omega}^*\bm{W}\bm{U}_{m,\omega}=\bm{V}_{m,\omega}^*\bm{W}\bm{V}_{m,\omega}=\bm{I},
\end{equation}
and recover the resolvent operator from equation \eqref{eqn:resolvent} as,
\begin{equation} \label{eq:svd}
\bm{R}_{m, \omega} = \bm{U}_{m,\omega} \bm{\Sigma}_{m, \omega} \bm{V}_{m, \omega}^* \bm{W}.
\end{equation}

For the resolvent analysis presented here, just as in \cite{schmidt2018spectral}, the above equations are discretized in the streamwise and radial directions with fourth-order summation by parts finite differences \citep{mattsson2004summation}, while the polar singularity is treated as in \cite{mohseni2000numerical} and non-reflecting boundary conditions are implemented at the domain boundaries.

\subsection{Spectral Proper Orthogonal Decomposition}

SPOD, similar to space-only proper orthogonal decomposition (POD) and originally shown by \cite{lumley1967, lumley1970},  determines an optimal (i.e. in terms of energy) set of orthogonal modes to describe a dataset, but unlike space-only POD, produces modes that express both spatial and temporal correlation in the data.  Like dynamic mode decomposition, SPOD modes are computed at unique frequencies.  However, through appropriate averaging, SPOD naturally ranks modes by energy and optimally accounts for the statistical variability of turbulent flows \citep{towne2018spectral}. Thus, the associated SPOD modes provide the ideal measurement tool to assess modes computed via resolvent analysis.

Decomposing the LES database $\bm{Q}$, where $\bm{Q}$ represents the temporal ensemble of perturbations ($\bm{q}^{\prime}$) found by applying the standard Reynolds decomposition, in the azimuthal and temporal dimensions via the discrete Fourier transform gives the decomposed data matrices, $\hat{\bm{Q}}_{m,\omega}$. Multiplying the decomposed matrices, at a particular frequency and azimuthal wavenumber, by their complex conjugate give the cross-spectral density
\begin{equation}
\textbf{S}_{m, \omega} = \hat{\bm{Q}}_{m,\omega} \hat{\bm{Q}}_{m,\omega}^*,
\end{equation}
to which we solve the SPOD eigenvalue problem presented by \cite{lumley1967, lumley1970}
\begin{equation}
\textbf{S}_{m, \omega}\textbf{W}\bm{ \Psi }_{m, \omega} = \bm{\Psi}_{m, \omega} \bm{\Lambda}_{m, \omega}. 
\end{equation}
The SPOD modes form the columns of $\boldsymbol{\Psi}_{m, \omega}$, ranked by the diagonal matrix of eigenvalues $\bm{\Lambda}_{m, \omega}= \text{diag}(\lambda_1, \lambda_2, ... , \lambda_N)$. The modes are orthonormal in the norm $\langle \cdot, \cdot \rangle_{E}$, and satisfy $\bm{\Psi}_{m, \omega}^*\bm{W}\bm{\Psi}_{m, \omega}= \bm{I}$. As a result, expansion of the cross-spectral density tensor gives, 
\begin{equation}
\textbf{S}_{m, \omega} = \bm{\Psi}_{m, \omega} \bm{\Lambda}_{m, \omega}\bm{ \Psi }_{m, \omega}^*. 
\end{equation}
In this study, we perform all SPOD computations with a Hamming window and realization sizes of 256 snapshots with 50\% overlap, resulting in 78 independent realizations.

To avoid ambiguity in referring to computed SPOD and resolvent modes, we use the following notation for the rest of the manuscript. First, all computed modes subscripts $m,\omega$ are dropped, but referenced when necessary in the text. Second, $\bm{\psi}_{n}$ represents the $n$-th most energetic SPOD mode, while $\bm{v}_n$ and $\bm{u}_n$ denote the resolvent forcing and response, respectively, that provide the $n$-th largest linear-amplification gain between $\bm{v}_n$ and $\bm{u}_n$. Finally, we use the notation $\bm{\psi}_{1}: u_x$ when referring to specific components of each mode, as shown here with streamwise velocity.

\subsection{Using SPOD to inform resolvent analysis} 
 \ethan{As SPOD provides the optimal description of the second order flow statistics, we wish to use this decomposition to inform our resolvent approach to match such statistics. The connection can be made through} multiplication of equation \eqref{eqn:resolvent} by its complex conjugate and then applying the expectation operator to present the relation between the CSD tensors of the forcing and response through the resolvent operator,
\begin{equation}
\boldsymbol{S}_{\bm{q}\bm{q}} = \mathbb{E} [ \bm{q}\bm{q}^* ] = \mathbb{E} [ \bm{R} \bm{f}\bm{f}^* \bm{R}^*] = \bm{R} \bm{S}_{\bm{f}\bm{f}}\bm{R}^*.
\end{equation}
If $\bm{q}$ is projected onto the SPOD modes and $\bm{f}$ is projected onto the input resolvent modes, $\bm{\beta} = \bm{V}^* \bm{W} \bm{f}$, where the vector $\boldsymbol\beta$ are the projection coefficients, then we may write 
\begin{equation}
\bm{\Psi} \bm{\Lambda} \bm{\Psi}^* = \bm{U} \bm{\Sigma} \bm{S}_{\boldsymbol\beta\boldsymbol\beta} \bm{\Sigma} \bm{U}^*,
\label{eqn:SPOD_res}
\end{equation}
which highlights that if the forcing coefficients are uncorrelated ($ \bm{S}_{\boldsymbol\beta\boldsymbol\beta} = \bm{\Lambda}_{\bm{\beta}}$ ) then the resolvent modes would be equivalent to the SPOD modes \citep{towne2018spectral}. Conversely, when the resolvent and SPOD modes are not identical, which is the case in our study, the forcing coefficients are correlated and this correlation must be modeled.

Rather than pursuing a direct model of the forcing coefficients, we take an alternative perspective that asks whether a modified resolvent operator, $\bm{R}_T$, can align one or more of the dominant resolvent and SPOD modes.  A trivial solution would be to define the operator by the SPOD expansion, i.e. ${\bm{R}}_T = \bm{\Psi}$, but this operator then corresponds to the (discretization of any) general (non-local) linear operator, rather than a specific partial differential equation (PDE).  Instead, a practical model can be obtained by posing a modified PDE of the linearized governing equations with one or more unknown coefficients, and then finding the best choice of coefficients such that the resolvent and SPOD modes are optimally aligned.  We propose such an approach in the next section by exploiting an eddy-viscosity model, and develop an optimization procedure that fits the parameters to align one, or more, of the most dominant resolvent and SPOD modes.

To the extent that the modified resolvent operator achieves \ethan{alignment of any one} of its output modes with a specific SPOD mode, we may directly interpret the corresponding diagonal entry of $\bm{S}_{\bm{\beta}\bm{\beta}}$ as the forcing amplitude, \ethan{ $\lambda_\beta$}, required to reproduce the SPOD mode amplitude $\lambda$, through the resolvent gain, $\sigma^2$.  In other words,
\begin{equation}
    \lambda_n = \sigma^2_n {\lambda_{\bm{\beta}}}_{n} \quad \mbox{for any $n$ where} \quad {\bm{u}}_n = {\bm{\psi}}_n,
    \label{eqn:betaI}
\end{equation}
independent of whether the other modes are aligned (as other modes are orthogonal).

\section{Models considered} \label{sec:models}
 
  \begin{table}
   \centering
\begin{tabular}{l c c c r}
 Turbulence model     \hspace{0.0cm} & $\bm{\mu}_T$  form  \hspace{0.0cm}                          & Optimal parameter  \hspace{0.0cm}    & LES data used    \hspace{0.0cm}        & Abbreviation        \\ 
 \hline
  Baseline*        & $1/Re_T = 3.\bar{3}\hspace{-0.09cm} \times \hspace{-0.09cm}10^{-5}$                            & --                     & --                        & Baseline \\
 Optimal field         & $\bm{\mu}_T(\bm{x})$                       & $\bm{\mu}_T(\bm{x})$   & $\bm{\Psi}$               &  Opt. $\bm{\mu}_T$  \\
 Mean-flow consistent  & $c \bm{\mu}_T(\bm{x})$                    & $c$                    & $\overline{\bm{q}}$                  & Mean $\bm{\mu}_T$   \\  
 RANS                  & $c \overline{\rho}  C_\mu k^2/\epsilon$   & $c$                    & --           & RANS $\bm{\mu}_T$   \\
 Turbulent $\text{Re}$        & $1/\text{Re}_T$                           & $1/\text{Re}_T$          & --                        & $\text{Re}_{T,Opt}$ \\
\end{tabular}
\caption{Turbulence models investigated in this study. The baseline* case refers to the results of \citet{schmidt2018spectral}.}
    \label{tab:models}
\end{table}


\ethan{We now add an eddy-viscosity model to the linearized governing equations~(\ref{eqn:LNS}). We follow the {\it ad hoc} model used in (amongst other references) \cite{del2006linear} and \cite{hwang2010linear}, which is typically justified by extending eddy viscosity from its traditional use in modeling the mean Reynolds stresses to modeling the effect of the ``background turbulence’’ on the coherent motion.}


The perturbation equations including the eddy viscosity are, with the replacement $\mu \mapsto \mu_j + \mu_T$, identical to the original linearized equations provided one accounts for the (spatial) variability of $\mu_T$ (equations provided in Appendix \ref{App:LNS}).  There remains an unknown forcing that is the residual between the original forcing and the ``coherent'' part that is is modeled by the eddy viscosity.  Unfortunately, the residual forcing no longer possesses its exact physical interpretation as the nonlinear interactions of resolved modes. However, the advantage is that the resulting response modes can significantly reduce the rank of the problem and lead to a residual forcing CSD that is tractable to model when compared to the forcing CSD of the exactly rearranged equations \citep{pickering2020resolvent,towne2020resolvent}.


In what follows, we refer to the modified linear operator with $\mu_T \ne 0$ as $\bm{A}_T$ and note that the operator depends on the chosen field for $\mu_T$, which, upon discretization becomes a vector $\bm{\mu}_T$.  Since we assume that $\mu_T$ is steady and axisymmetric, the operators have a similar temporal/azimuthal Fourier transform that we denote ${\bm{A}_{T}}_m$.

We now consider four models for the eddy-viscosity field. The first model directly optimizes the eddy-viscosity field to maximize alignment between the dominant resolvent and SPOD modes.  The second model fits an eddy viscosity to the LES mean flow by minimizing the residual in the steady RANS equations.  The third model uses an independently computed eddy-viscosity field from a RANS $k-\epsilon$ model.  Finally, we consider a simpler constant eddy-viscosity model based solely upon a turbulent Reynolds number. 

For brevity, we refer to the modes computed with the above eddy-viscosity models as EVRA (eddy-viscosity resolvent analysis) modes, while modes termed ``baseline’’ refer to those computed by \cite{schmidt2018spectral}. We chose this study as reference for its extensive comparison of resolvent and SPOD modes across all three turbulent jets and many wavenumbers and frequencies. In the baseline study, they chose an effective Reynolds number of $Re_T = 3 \times 10^4$, a value that is an order of magnitude smaller than the molecular Reynolds number, yet not consistent with the expected magnitude of an eddy viscosity (i.e. $Re_T<<  3 \times 10^4$) .  Instead, we regard this intermediate value as a regularization of the resolvent operator. Table \ref{tab:models} summarizes the various models investigated.

For exploratory purposes, we find an eddy-viscosity field that best aligns the (so modified) resolvent operator to the measured SPOD modes independently for each frequency and azimuthal mode. The purpose is to gauge the sensitivity of the eddy viscosity value needed to model the different frequencies and azimuthal modes, and should not be interpreted as a proposal for a frequency-dependent eddy viscosity.

Parenthetically, within the following optimization framework we can consider any turbulence model or regularization based on mean-flow quantities.  A further example is given in appendix \ref{App:Linear}, where we consider a linear damping model recently proposed for resolvent analysis of unstable base flows \citep{yeh2019resolvent}.

\subsection{Optimal eddy-viscosity field} \label{sec:Opt_eddy}

Here we develop an optimization, computed independently for each frequency and azimuthal mode, that finds the eddy-viscosity field that is optimal (i.e. the upper bound) in aligning the leading resolvent and SPOD modes.  To find the analytical expression that determines the sensitivity of mode alignment to an eddy-viscosity field, we use a Lagrangian technique analogous to \cite{brandt2011effect} that accounts for the non-modal behavior of the resolvent operator. This technique couples constraints from the governing equations, resolvent analysis, a normalization, and a cost function (alignment of leading SPOD and resolvent modes), into a Lagrangian functional for whose stationary point provides the desired maximum.

To build the Lagrangian functional, we begin with the forward equation \eqref{eqn:LNS_eq} and substitute $\bm{L}$ with $\bm{L}_T$, the linear operator that includes an eddy-viscosity model. The singular value / singular vector $ (\bm{v}_1,\bm{u}_1, \sigma_1) $ as defined in \eqref{eq:svd} is a solution of  both the forward equation \eqref{eqn:LNS_eq}, 
 \begin{align}
 \bm{v}_1 &= \bm{L}_T \bm{u}_1 \label{eqn:res_eig0},
 \end{align}
 where $\bm{v}_1$ replaces $\bm{f}$ as the forcing and $\bm{u}_1$ replaces $\bm{q}$ as the associated response, and the resolvent eigenvalue problem,
 \begin{align}
 \bm{W} \bm{u}_1 &= \sigma_1^2 \bm{L}_T^* \bm{W}\bm{v}_1.
 \label{eqn:res_eig}
 \end{align}
The above resolvent eigenvalue solution is found by taking the energy norm of equation \eqref{eqn:resolvent} and dividing by the forcing energy to give
 \begin{equation}
     \frac{\bm{u}_1^* \bm{W} \bm{u}_1}{\bm{v}_1^* \bm{W} \bm{v}_1} = \sigma_1^2 = \frac{\bm{v}_1^* \bm{R}_T^* \bm{W} \bm{R}_T \bm{v}_1}{\bm{v}_1^* \bm{W} \bm{v}_1}.
 \end{equation}
 Rearranging and eliminating $\bm{v}^*_n$ we arrive at  
 \begin{equation}
     \bm{R}_T^* \bm{W} \bm{R}_T \bm{v}_1 = \sigma_1^2 \bm{W} \bm{v}_1, 
 \end{equation}
 where replacing $\bm{R}_T \bm{v}_1$ with $\bm{u}_1$ and multiplying both sides by $\bm{R}^{-*}_T = \bm{L}_T^*$ recovers equation \eqref{eqn:res_eig}. Finally,we define a normalization constraint via,
    \begin{equation}
     \langle \bm{u}_1, \bm{u}_1 \rangle_{E} = \bm{u}_1^* \bm{W} \bm{u}_1 = 1.
     \label{eqn:constraint}
     \end{equation}
 
 The last component of the Lagrangian functional is the cost function, 
 \begin{equation}
         \mathcal{J} = \bm{u}_1^* \bm{W} \bm{\psi}_1 \bm{\psi}_1^* \bm{W} \bm{u}_1-l^2\bm{\mu}_T^*\bm{M}{\bm{\mu}_T},
         \label{eqn:cost}
 \end{equation}
where the first term, representing the primary objective, measures the squared projection, or alignment, between the dominant SPOD mode, $\bm{\psi}_1$, and first resolvent mode, $\bm{u}_1$. The alignment measure, $\bm{u}_1^* \bm{W} \bm{\psi}_1$, is squared to ensure the cost function is real. For brevity, we denote the outer product of the dominant SPOD mode as $\bm{\Psi}_1 = \bm{\psi}_1\bm{\psi}_1^* = \bm{\Psi}_1^*$. The cost function may also consider multiple resolvent / SPOD modes by considering a (weighted if desired) sum of the squared alignment terms. 

The second term, $-l^2 \bm{\mu}_T^* \bm{M} {\bm{\mu}_T}$, is a Tikhonov regularization that penalizes values of ${\bm{\mu}_T} $ that do not affect the alignment (high values of ${\bm{\mu}_T} $ diminish the value of $ \mathcal{J}$), with $\bm{M}$ representing the cylindrical quadrature weights of the grid. As done in standard regularization methods, the value of $ l^2$ is chosen high enough to remove the values of ${{\bm{\mu}_T}}$ in insensitive regions, but also sufficiently small to not interfere with the primary objective \citep{Hansen1993}. This penalization is effective at minimizing the eddy viscosity in  non-turbulent regions of the flow such as the far field.  A substantial range of $l^2$ values (i.e. multiple orders of magnitude) remove negligible regions of the eddy-viscosity field from the initial field without an observable drop in the primary objective, alignment between $\bm{u}_1$ and $\bm{\psi}_1$.

We now formally construct the Lagrangian functional to include the cost function \eqref{eqn:cost}, forward equation \eqref{eqn:res_eig0}, the resolvent eigenvalue problem \eqref{eqn:res_eig}, and the normalization constraint \eqref{eqn:constraint} to give,
     \begin{align}
     \mathcal{L} &= \bm{u}_1^* \bm{W} \bm{\Psi}_1 \bm{W} \bm{u}_1-l^2\bm{\mu}_T^*\bm{M}{\bm{\mu}_T} \nonumber \\ &- \tilde{\bm{u}}_1^* (\bm{L}_T \bm{u}_1 - \bm{v}_1 ) 
     - \tilde{\bm{v}}_1^*  (  \bm{W}\bm{u}_1 -\sigma_1^2 \bm{L}_T^*\bm{W} \bm{v}_1 )  
     - \tilde{\sigma}_1 ( \bm{u}_1^* \bm{W} \bm{u}_1 - 1  ) +c.c.,
     \end{align}
where $( \tilde{\bm{u}}_1, \tilde{\bm{v}}_1, \tilde{\sigma}_1)$ are Lagrange multipliers and $\tilde{\sigma}_1$ is real-valued as the corresponding constraint is real).  This results in a functional that depends on seven variables,
     \begin{equation}
     \mathcal{L}  ( [ \bm{u}_1, \bm{v}_1 , \sigma_1], [ \tilde{\bm{u}}_1 , \tilde{\bm{v}}_1, \tilde{\sigma}_1 ], \bm{\mu}_T).
     \end{equation}
     
We can find the maximum of the cost function by finding the stationary point of the entire functional (i.e. where variations with respect to each variable are zero). Stationarity with respect to the Lagrange multipliers yields the state equations, which are by definition satisfied, while stationarity  with respect to the state variables yields:
\begin{align}
     & \frac{\partial \mathcal{L}}{\partial \bm{u}_1} \delta \bm{u}_1 =   ( 2 \bm{W} \bm{\Psi}_1 \bm{W} \bm{u}_1 - \bm{L}_T^* \tilde{\bm{u}}_1 - \bm{W} \tilde{\bm{v}}_1 - 2 \tilde{\sigma}_1 \bm{W} \bm{u}_1)^* \delta \bm{u}_1 &&= 0\\
     &\frac{\partial \mathcal{L}}{\partial \bm{v}_1} \delta \bm{v}_1= (\tilde{\bm{u}}_1 + \sigma_1^2  \bm{W} \bm{L}_T \tilde{\bm{v}}_1)^*\delta \bm{v}_1 && = 0 \\
    & \frac{\partial \mathcal{L}}{\partial \sigma_1} \delta {\sigma}_1= (\tilde{\bm{v}}_1^* \bm{L}_T^* \bm{W} \bm{v}_1)^* \delta \sigma_1 && = 0, 
\end{align}
and the condition in the last equation may be simplified into $\tilde{\bm{v}}_1^* \bm{L}_T^* \bm{W} \bm{v}_1 = \tilde{\bm{v}}_1^* \bm{W}{\bm{u}}_1$ using  equation \eqref{eqn:res_eig}. The stationary point is subsequently met by constructing the following system of equations and solving for the Lagrange multipliers:
\begin{align}
      \begin{bmatrix}
     - \bm{L}_T^* & -\boldsymbol{W} & - 2 \bm{W} \bm{u}_1 \\
     \boldsymbol{W}^{-1} & \bm{L}_T \sigma_1^2  & 0 \\
     0 & \bm{u}_1^* \bm{W} & 0
     \end{bmatrix}
     \begin{bmatrix} \tilde{\bm{u}_1}\\ \tilde{\bm{v}_1} \\ \tilde{\sigma}_1
     \end{bmatrix} 
     & =       
     \begin{bmatrix} - 2 \bm{W} \bm{\Psi}_1 \bm{W} \bm{u}_1 \\ 0 \\ 0 \end{bmatrix}. \label{eq:syst}
\end{align}
The upper left $ 2\times 2$ block is degenerate due to the state equations
\eqref{eqn:res_eig0} and \eqref{eqn:res_eig} (the couple, $\tilde{\bm{u}}_1=\bm{W}\bm{v}_1$ and $\tilde{\bm{v}}_1=-\sigma_1^{-2}\bm{u}_1 $, is in the null-space of this block) and the third column and line regularizes this system. Combining the 3 equations, one can show that $\tilde{\sigma}_1= \bm{u}_1^*\bm{W} \bm{\Psi}_1 \bm{W} \bm{u}_1$, proving that $\tilde{\sigma}_1$ is a real value.

\begin{algorithm}[t!]
  \caption{Optimization}\label{algorith1}
  \begin{algorithmic}[1]
\setstretch{1.0}
    \State 
    Initialize. Choose an initial eddy-viscosity/turbulence model and target SPOD mode.

    \WHILE{$d\mathcal{J} / d \bm{\mu}_T \neq \bm{0}$}

        \STATE
        Compute the EVRA mode(s).
        
        \STATE
        Solve for the Lagrange multipliers.
        
        \STATE
        Calculate the update direction, $d\mathcal{J} / d \bm{\mu}_T$.
        
        \STATE
        Determine the optimal value of the step $ \alpha $ by repeated evaluation of the cost functional along the steepest ascent direction.

    \ENDWHILE
  \end{algorithmic}
\end{algorithm}

\begin{figure}
\centering
\includegraphics[width=1\textwidth,trim={0cm 0cm 0cm 0cm},clip]{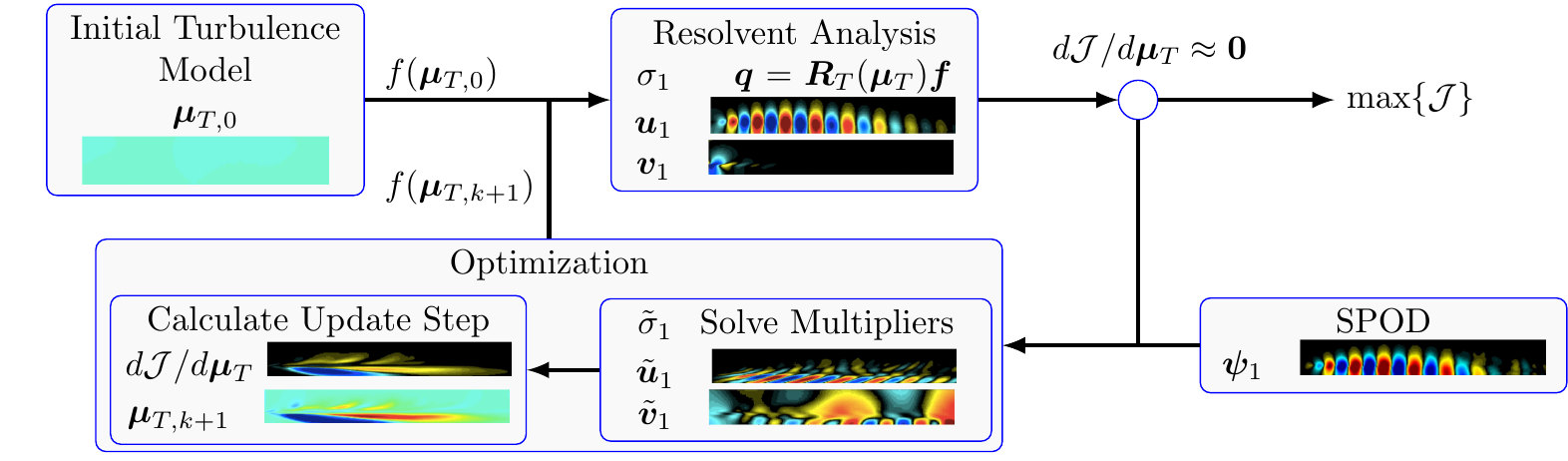}
\caption{Schematic of the optimization framework for determining the optimal eddy-viscosity field that maximizes the alignment between computed resolvent modes, $\bm{u}_1$, and educed SPOD modes, $\bm{\psi_1}$. Included graphics are from implementation of the full-field eddy-viscosity model at $St=0.6$, $m=0$, and $M_j=0.4$.}
\label{fig:Optimization_Schematic}
\end{figure}

A final variation is taken with respect to the eddy-viscosity, ${\bm{\mu}_T}$ (which may be a scalar or vector quantity), providing the direction of gradient ascent for the eddy-viscosity field,
\begin{align}
        \frac{\partial \mathcal{L}}{\partial {\bm{\mu}_T}}{\delta {\bm{\mu}_T}} &=  - \tilde{\bm{u}}_1^* \left( \frac{\partial \bm{L}_T}{\partial {\bm{\mu}_T}} {\delta  {\bm{\mu}_T}}\right)\bm{u}_1 + \sigma_1^2 \tilde{\bm{v}}_1^* \left(  \frac{\partial \bm{L}_T^*}{\partial {\bm{\mu}_T}} {\delta {\bm{\mu}_T}}\right) \bm{W}  \bm{v}_1 - 2l^2 \bm{\mu}_T^* \bm{M} {\delta {\bm{\mu}_T}}+c.c \\
        &=\left(\frac{d\mathcal{J}}{d{\bm{\mu}_T}}\right)^*\bm{M}\delta{\bm{\mu}_T}.
\end{align}
The gradient at the $k^{th}$ grid point is then:
\begin{equation} \label{eq:grad}
    \left.\frac{d{\mathcal J}}{d\bm{\mu}_T}\right|_k=\bm{M}_{km}^{-1}\left(-\bm{u}_{1,j}^* \bm{L}_{m,ij}^*\tilde{\bm{u}}_{1,i}+\sigma_1^2\bm{W}_{lj}\bm{v}_{1,l}^* \bm{L}_{m,ji}\tilde{\bm{v}}_{1,i}\right)-2l^2 \bm{ \mu}_{T,k}+c.c,
\end{equation}
where $ \bm{L}_{m,ij} = \mbox{lim}_{\epsilon \rightarrow 0} \frac{\bm{L}_{T+\epsilon \delta\bm{\mu}_{m},ij}-\bm{L}_{T,ij}}{\epsilon}$, $\delta\bm{\mu}_{m}$ being a null vector except at the $m^{th}$ position where it is equal to 1. This tensor may be obtained either through automatic differentiation of $\bm{L}_T$ with respect to ${\bm{\mu}_T}$ or by finite differences. Full storage of such tensors is not an issue when finite differences, finite volumes, or finite elements are used for the spatial discretization as the resulting tensors are extremely sparse. 

The updated optimization parameter is then:
\begin{align}
    \bm{\mu}_T^{(k+1)} = \bm{\mu}_T^{(k)} + \alpha \frac{d\mathcal{J}}{d\bm{\mu}_T},
\end{align}
where $k$ is the iteration number and $\alpha$ is a step size determined through a root finding algorithm or a line search. If multiple SPOD /resolvent modes are considered for the optimization then one has to solve equation \eqref{eq:syst} for each couple $ [\bm{\Psi}_n, (\bm{v}_n,\bm{u}_n, \sigma_n) ]$ and the total gradient $\frac{d\mathcal{J}}{d\bm{\mu}_T}$ is the sum of each individual gradient, while the line search for $ \alpha $ is performed considering the full cost functional. Although considering multiple modes is theoretically straightforward (and we present one example in \S~\ref{sec:subdominant}), there are two practical issues. Each additional mode brings further complexity to the gradient, increasing computation time, and the quality of SPOD modes, $\bm{\Psi}_n$, become increasingly noisy with $n$, thus rendering gains via the optimization as marginal. We discuss the latter issue in more detail throughout the manuscript. Figure \ref{fig:Optimization_Schematic} presents a schematic of the above optimization framework, including graphical examples from the optimal eddy-viscosity field case at $St = 0.6$, $m=0$, and $M_j = 0.4$.

For some cases, the optimization step imparts a region of negative eddy viscosity presenting a challenge in both its physical interpretation and the numerical stability of the resolvent operator. However, negative eddy viscosity is not a unique concept to the algorithm presented. Literature surrounding eddy-viscosity models used in RANS and LES attribute physical interpretations of negative eddy-viscosity to backscattering of turbulent energy, which, in many simulations, results in unstable simulations \citep{ghosal1995dynamic}. Common treatment of a negative eddy viscosity has included filtering operations, ensemble averaging in homogeneous directions, and ad hoc clipping of the eddy-viscosity field \citep{vreman2004eddy}, while inferences of the eddy-viscosity field via a Boussinesq approximations of data are often regularized to remove negative regions (e.g. \cite{semeraro2016modeling}). Here, we also elect to remove any negative eddy viscosity using a simple clipping strategy by setting any negative regions to zero such that only the molecular viscosity is present.

The topology of the proposed cost function is complex, as $\bm{\mu}_T$ involves many degrees of freedom, and our optimizer may return a local rather than global maximum. 
Therefore, a complete assessment of the sensitivity of initial conditions or demonstration of a global maximum are intractable, but the relative insensitivity of the results to initial guesses and the fact that no other considered method outperforms the full optimization (shown later in figure \ref{fig:alignments}) provide confidence in the robustness of the maxima achieved. For all of the results presented here, we use the optimal constant eddy-viscosity field results (introduced in \S~\ref{sec:Constant}) as the initial condition for the full-field optimizations.

\Revthree{Finally, the above optimization is derived considering the full (perturbation) state as the output. The formulation is similar if the input and output spaces are restricted, as shown in Appendix \ref{App:inputoutput}.}


\subsection{Mean-flow consistent eddy-viscosity model}\label{sec:Opt_mean_eddy}

For many experimental and numerical datasets, including the LES databases used here, an eddy-viscosity field is absent. We circumvent this issue by finding the eddy-viscosity field that minimizes the error to which the mean flow satisfies the (zero frequency and axisymetric wavenumber) linearized Navier-Stokes equations, supplemented with an eddy-viscosity model, provided in Appendix \ref{App:LNS}. To do so, we find an eddy-viscosity field that minimizes the residual $\overline{\bm{f}}$ given by
\begin{equation}
    \bm{L}_T\overline{\bm{q}} = \overline{\bm{f}}.
\end{equation}
Thus we define the cost function,
  \begin{equation}
      \mathcal{J} = -\overline{\bm{f}}^* \bm{W} \bm{\overline{\bm{f}}},
  \end{equation}
and develop a Lagrangian  functional with the forward equation as the only additional constraint to give
  \begin{equation}
      \mathcal{L} = -\overline{\bm{f}}^* \bm{W} \bm{\overline{\bm{f}}} - \tilde{\bm{u}}^* (\bm{L}_T\overline{\bm{q}} - \overline{\bm{f}}).
  \end{equation}
Variations with respect to the residual are
\begin{equation}
    \frac{\partial \mathcal{L}}{\partial \overline{\bm{f}}}\delta\overline{\bm{f}} = (- 2 \bm{W} \overline{\bm{f}} + \tilde{\bm{u}})^*\delta\overline{\bm{f}} = 0,
\end{equation}
and we may directly solve for the Lagrange multipliers as,
\begin{equation}
     \tilde{\bm{u}}= - 2\bm{W} \overline{\bm{f}}.
\end{equation}
Then by taking variations with respect to the eddy-viscosity field gives,
\begin{equation}
    \frac{\partial \mathcal{L}}{\partial \bm{\mu}_T} \delta \bm{\mu}_T= - 2 (\bm{W}\overline{\bm{f}})^*  \left( \frac{\partial \bm{L}_T}{\partial \bm{\mu}_T}\delta\bm{\mu}_T\right)\overline{\bm{q}}.
\end{equation}
Similar to equation \eqref{eq:grad}, we obtain the update step:
\begin{equation}
    \left.\frac{d{\mathcal J}}{d\bm{\mu}_T}\right|_k=-2\bm{M}_{km}^{-1}\overline{\bm{q}}_{j} \bm{L}_{m,ij}\bm{W}_{il} \overline{\bm{f}}_{l},
\end{equation}
and find the field via a line search. These steps are described in greater detail in the preceding subsection \S~\ref{sec:Opt_eddy} . Figure \ref{fig:Eddy_fields}~(a) provides the eddy-viscosity field that optimally minimizes the residual of the mean-flow solution. The associated residual field for this model reduced errors to approximately 10\% of the original residual field, with the exception where the shear layer is thin near the nozzle. The thin shear-layer region improved by only $\approx$ 50\%, but as shown later in the manuscript, modes in this region are generally less sensitive to the eddy-viscosity field. 

We refer to this model as the mean-flow consistent eddy-viscosity model and we optimally tune this field at each frequency by introducing the coefficient, $c$, $\bm{\mu}_T = c \bm{\mu}_{T,Mean}$. Our interest in the value of $c$ is not to propose a functional of its frequency dependence (or assign to it a physical meaning), but to measure and observe the overall variation and help determine whether a frequency independent coefficient might suffice.
 
\subsection{RANS-based eddy-viscosity field} \label{sec:RANS}

We compute steady-state RANS solutions for each case to assess the applicability of the associated eddy-viscosity field for resolvent analysis. For simplicity, we perform the RANS computations in Fluent. The 2D axisymmetric grid extends 40 diameters in the streamwise directions and 20 diameters in the radial direction with grid spacing mirroring that of the interpolated LES grid scaled to be four times finer, giving $3 \times 10^5$ grid points. We set the inlet boundary conditions to the base-flow profile from the LES simulations and use the standard 2-equation $k-\epsilon$ model \citep{launder1983numerical} for turbulence modeling. Coefficients used for the model are variants of those suggested by \cite{thies1996computation}, with turbulent viscosity coefficient $C_\mu = 0.0874$, dissipation transport coefficients $C_{\epsilon1} = 1.4$ and $C_{\epsilon2} = 2.02$, turbulent Prandtl numbers for kinetic energy $\sigma_k = 0.324$ and dissipation $\sigma_\epsilon = 0.377$, and the turbulent Prandtl number $Pr_T = 0.422$. However, the standard $\kappa-\epsilon$ model provided in ANSYS does not incorporate the \cite{pope1978explanation} and \cite{sarkar1991analysis} correction terms used in \cite{thies1996computation}, requiring a calibration of the mean-flow quantities by introducing a scaling constant $a$ to $C_\mu = 0.0874 / a$, $\sigma_K = 0.324 / a$, and $\sigma_\epsilon = 0.377 / a$.

\begin{figure}
(a) \\  \includegraphics[width=1\textwidth,trim={1cm 1.35cm 0cm 1cm},clip]{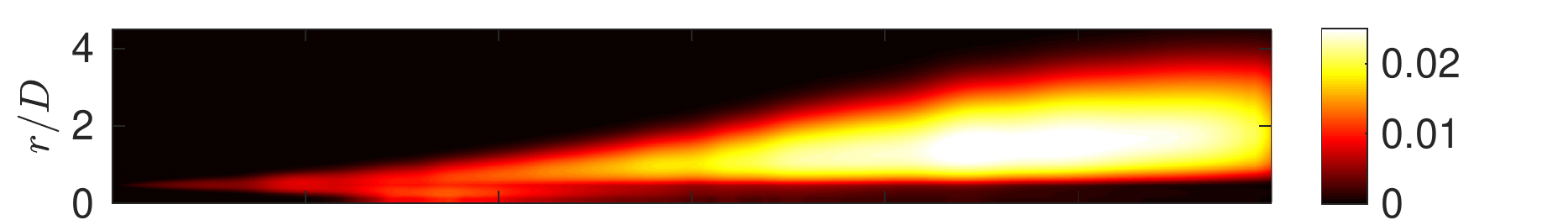}
(b) \\ \includegraphics[width=1\textwidth,trim={1cm 0cm 0cm 1cm},clip]{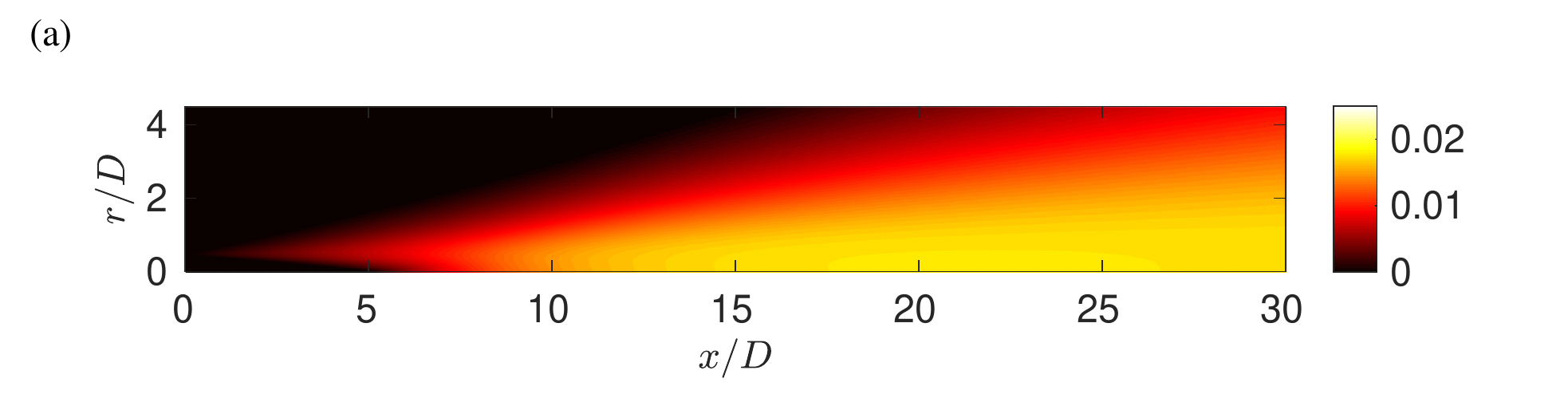}
\caption{ (a) Mean-flow consistent eddy-viscosity model computed at zero frequency and azimuthal wavenumber. (b) Eddy-viscosity field computed via a RANS simulation for the $M_j = 0.4$ jet, $c=1$. }
\label{fig:Eddy_fields}
\end{figure}

\begin{figure}
\hspace{0.5cm} (a) \hspace{5.5cm} (b) \\
\includegraphics[width=1\textwidth]{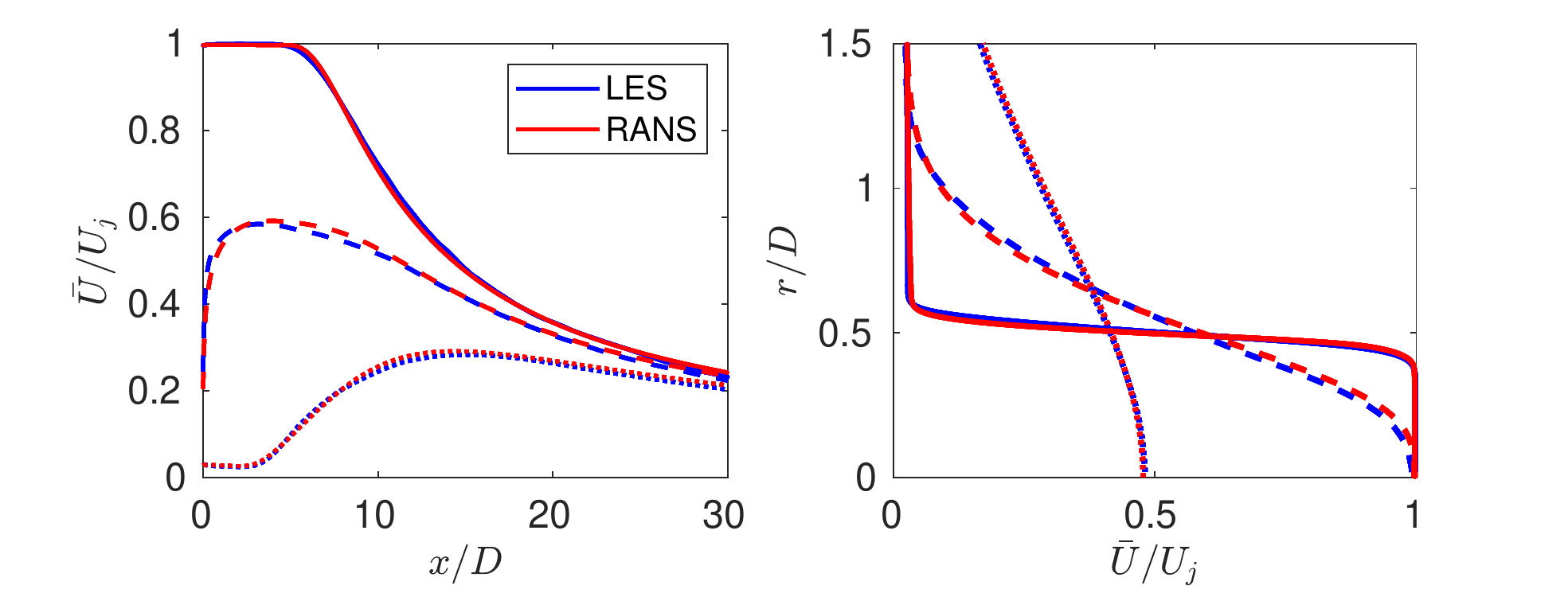}
\caption{Mean-flow profiles of both the $M_j = 0.4$ LES and RANS, where the RANS simulation was tuned to best match the LES mean flow. (a) presents the streamwise mean velocity at three radial locations, $r/D =$ {\solidrule} 0.25, {\dashedrule} 0.5, {\dottedrule} 1, versus streamwise distance from the nozzle, while (b) gives the streamwise mean velocity at three streamwise locations, $x/D =$ {\solidrule} 0.5, {\dashedrule} 5, {\dottedrule} 10, versus radial distance.}
\label{fig:M04_RANS_Baseflow}
\end{figure}

RANS mean-flow quantities closely match those of the LES for each of the three turbulent jets using values for $a$ of 1.2, 1.3, and 1.575, for $M_j = 0.4$, $0.9$, and 1.5, respectively. While tuning of the constant $a$ to match LES in not in the spirit of obtaining a universal RANS model, we do so here to give the RANS-generated eddy-viscosity field the best chance at being consistent with the LES results from which the SPOD modes were educed.  For a full assessment of the accuracy of RANS predictions for turbulent jets we refer the reader to \cite{thies1996computation,georgiadis2006evaluation}.

Figure \ref{fig:Eddy_fields}~(b) presents the RANS-predicted eddy-viscosity field for the $M_j = 0.4$ jet and figure \ref{fig:M04_RANS_Baseflow} shows near identical agreement with the mean LES streamwise flow.  We observe similar agreement in radial velocity, density, and turbulent kinetic energy, and also find close agreement for the $M_j=0.9$ and 1.5 jets; we do not show these results for brevity. For determination of the optimal RANS-based eddy-viscosity field at each frequency, we take the computed eddy-viscosity fields, 
\begin{align}
         \bm{\mu}_{T,RANS} =  \overline{\rho}  C_\mu \frac{k^2}{\epsilon},
\end{align}
and introduce the coefficient, $c$, $\bm{\mu}_T = c \bm{\mu}_{T,RANS}$ (just as in \S~\ref{sec:Opt_mean_eddy}).
    
\subsection{Constant eddy-viscosity field} \label{sec:Constant}

Finally, we consider a simple, constant eddy viscosity, $\bm{\mu}_T = 1/ Re_T$. We primarily investigate this model because of its use in many turbulent jet studies that used a Reynolds number based either upon the molecular viscosity \citep{jeun2016input,lesshafft2019resolvent}, on the order of $10^5-10^6$, or through an effective turbulent viscosity \citep{garnaud2013preferred,schmidt2018spectral}, on the order of $10^3-10^4$. These, quite different, choices inevitably provided discrepancies in amplification gains and mode shapes across each study, particularly at low frequencies (i.e. $St < 0.3$ for $m=0$) – showing that the Reynolds stresses have a substantial impact on resolvent analyses of turbulent jets.  Here, we find the optimal $\text{Re}_T$ at each frequency and azimuthal mode number by a line search.
 
\section{Optimal SPOD and resolvent mode alignment} \label{sec:results}

\begin{figure}
\hspace{0.35cm} (a) \hspace{5.675cm} (b) \\
	\includegraphics[width=0.49\textwidth]{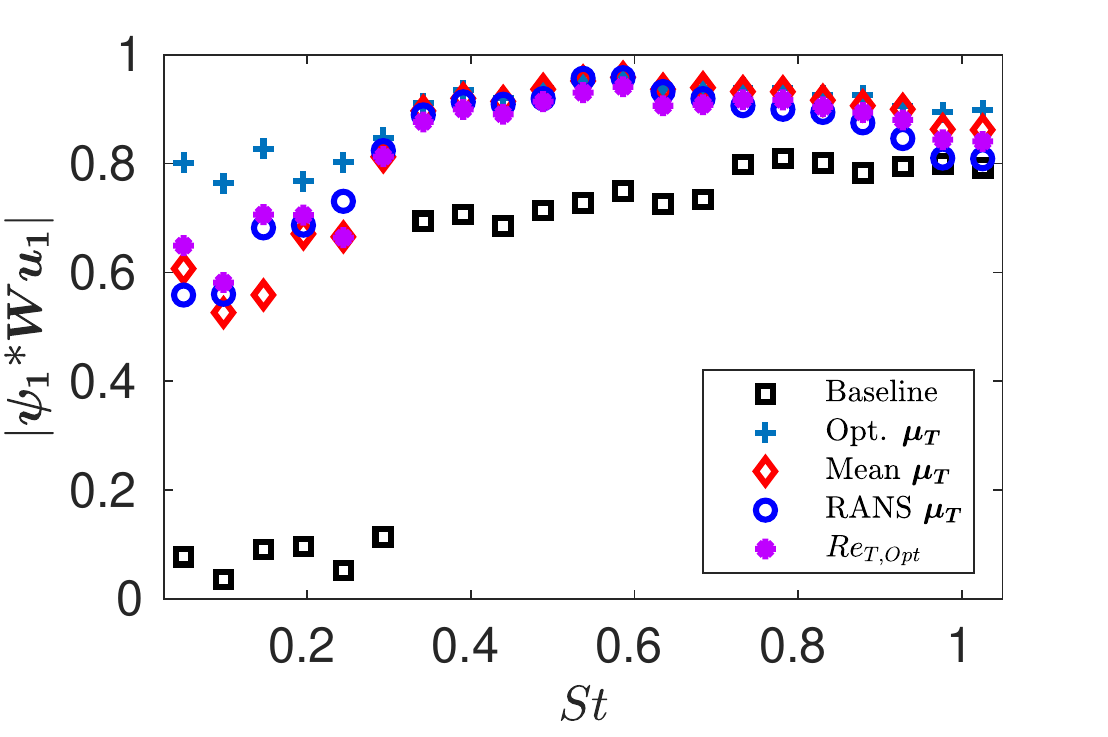}
	\includegraphics[width=0.49\textwidth]{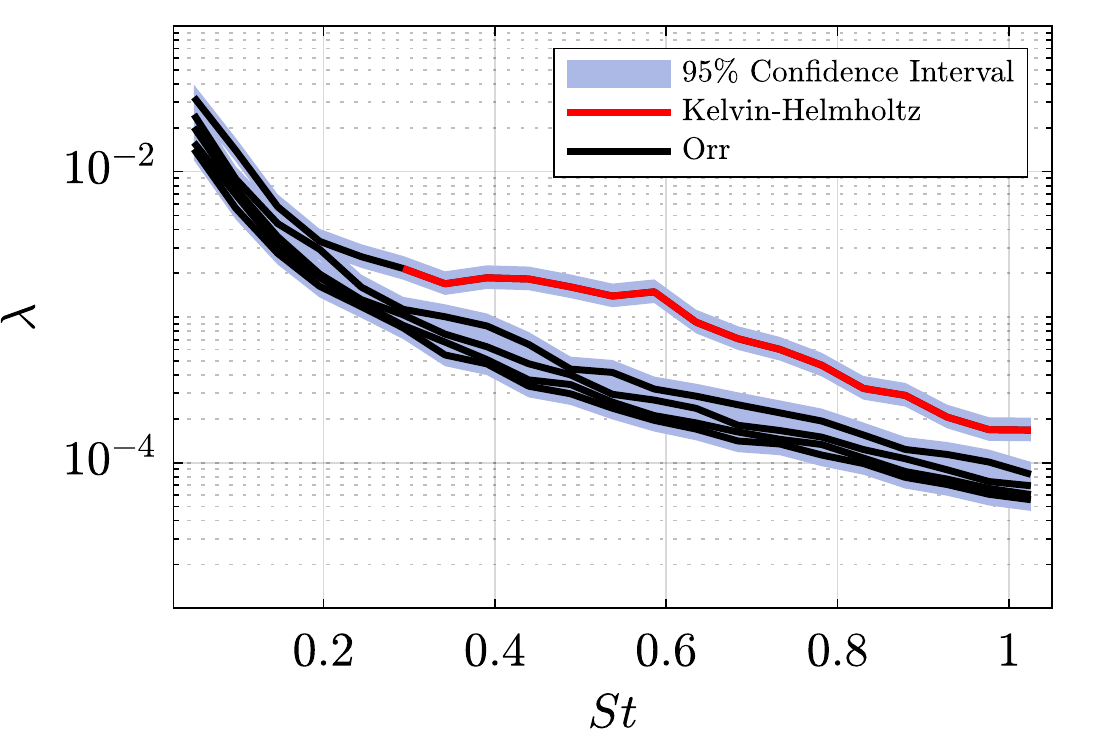}
	\caption{(a) Optimal alignments for all methods investigated including the baseline case, $Re_T = 3 \times 10^4$. (b) SPOD eigenvalue spectra of the first five modes for $m=0$, including the 95\% confidence intervals and the modes associated with the Kelvin-Helmholtz and Orr mechanisms.}
	\label{fig:alignments}
\end{figure}

In this section, we present modes predicted by the various EVRA models presented in the previous section. We focus on the axisymmetric disturbances, $m=0$, for the $M_j=0.4$ jet, and report results for other azimuthal modes and jet Mach numbers in section~\ref{sec:complete}. We performed optimizations over the frequency range $St \in[0.05,1]$, resulting in the alignment coefficients displayed in figure \ref{fig:alignments}, with alignment defined as $|\bm{\psi}_1^* \bm{W} \bm{u}_1|$. This metric not only represents how similar the spatial structures, represented as complex eigenfunctions, are between the dominant resolvent and SPOD modes, but also measures the similarity in distribution of energy amongst the five state variables.  A value of 1 signifies perfect agreement, giving both identical agreement in structure and distribution of energy in the state variables.  Typically, in this metric, values of approximately 0.4 or greater show qualitative agreement, whereas values less than 0.4 have little visual similarity. 

Figure \ref{fig:alignments}~(a) shows that throughout the frequency range considered, the alignments improve considerably from the baseline case (constant eddy viscosity with $Re_T = 3 \times 10^4$). The alignment is best for $St >0.3$, which corresponds to the frequencies where the jet has a strong, low-rank Kelvin-Helmholtz (KH) response \citep{schmidt2018spectral}, as highlighted by figure \ref{fig:alignments}~(b), presenting the spectra of the first five SPOD modes and their 95\% confidence interval. For this region, $St>0.3$, the baseline case gives reasonable (> 75\% alignment) results, nonetheless, the eddy-viscosity models still improve the modes to nearly perfect alignment. At lower frequencies, $St \leq 0.3$, we find the most dramatic increase in alignments, from approximately 10\% to 80\%. These substantial improvements, at $St \leq 0.3$, coincide with a change of mode type, from KH to Orr \citep{schmidt2018spectral}, a viscous, non-modal instability mechanism sensitive to Reynolds number (with rapidly increasing amplification as Reynolds number increases), that dominates the non-optimized, low-frequency and subdominant regions of the resolvent spectrum for the $M_j = 0.4$ jet.  We also find that the optimal eddy-viscosity field provides the greatest alignment among the models, which is at least suggestive that the optimization achieved a global maximum. 

Surprisingly, the other eddy-viscosity models produce alignments close to the optimal eddy-viscosity field. The constant eddy-viscosity is nearly optimal at lower frequencies (Orr-type modes), whereas the RANS and optimal mean-flow eddy-viscosity models are more nearly optimal at higher ones. We stress that in the optimal mean-flow, RANS, and constant $\bm{\mu}_T$ models, a different optimal value of the coefficient (i.e. $c$ and $Re_T$) is used at each frequency. We differ a discussion of the sensitivity of these coefficients to \S~\ref{sec:Sensitivity}.

\begin{figure}
\centering
\vspace{0.5cm}
\includegraphics[width=1\textwidth]{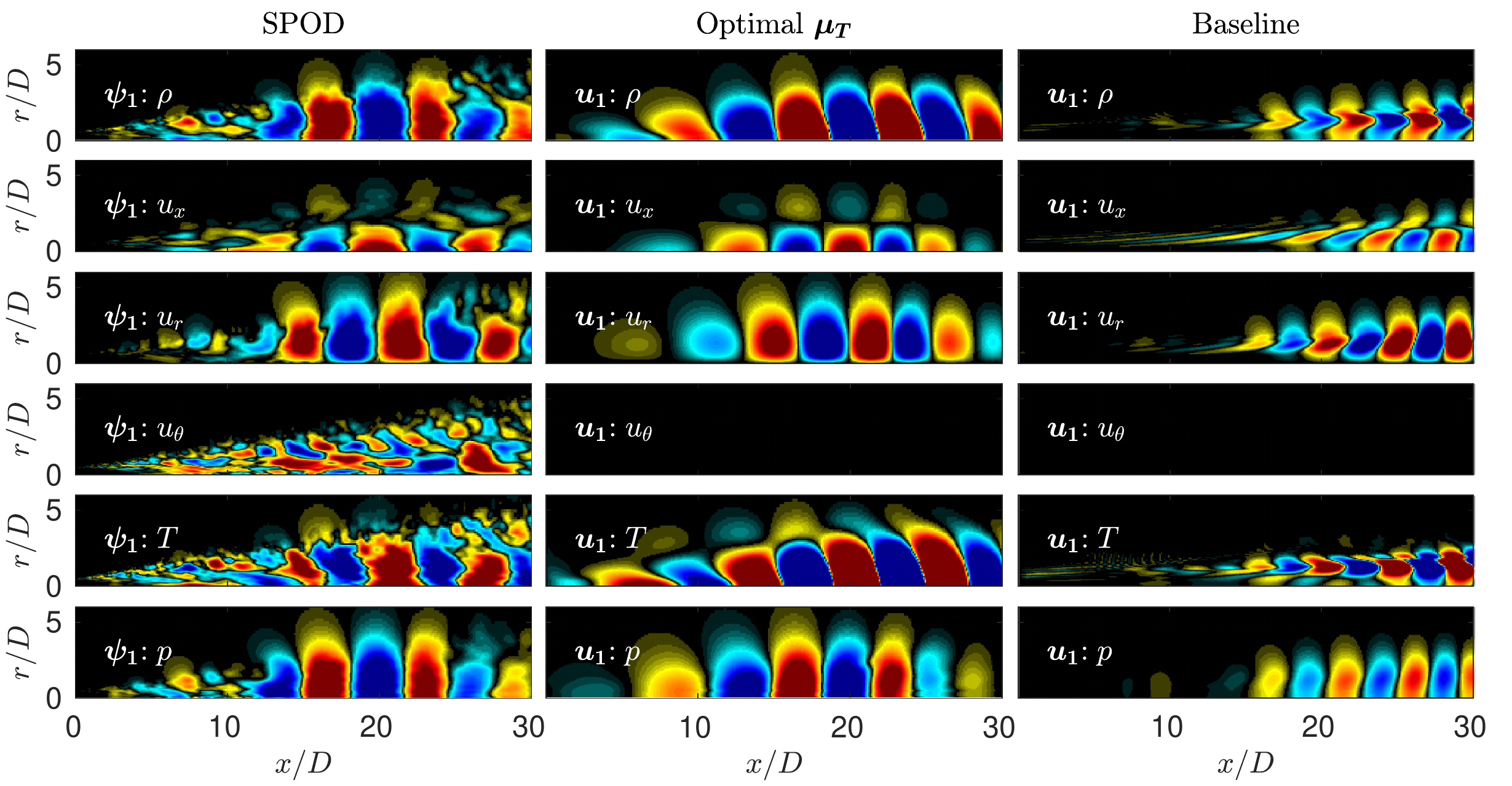}
\caption{Real component of the fluctuating response state variables, $\bm{q}^{\prime}= [\rho, u_x, u_r, u_\theta, T]$, and pressure, $p$, at $St = 0.05$, $m=0$. The columns display SPOD $(\bm{\psi}_1)$, optimal eddy viscosity ($\bm{u}_1$), and baseline ($\bm{u}_1$) modes from left to right, respectively. Contours (${\color{Red}\blacksquare}\!{\color{Black}\blacksquare}\!{\color{Blue}\blacksquare}$) are given by $\pm 0.5||\bm{\psi}_1: \cdot ||_\infty$ of the SPOD mode, where $\cdot$ is the fluctuating variable in question (with $||\bm{\psi}_1:\cdot||_\infty$ values: $[\rho, u_x, u_r, u_\theta, T, p] =
[2.8, 198.6, 46.0, 37.2, 1.2, 10.4] \times 10^{−3}$).}
\label{fig:St005}
\end{figure}

\begin{figure}
\centering
\vspace{0.5cm}
\includegraphics[width=1\textwidth,trim={0cm 13.125cm 0cm 0cm},clip]{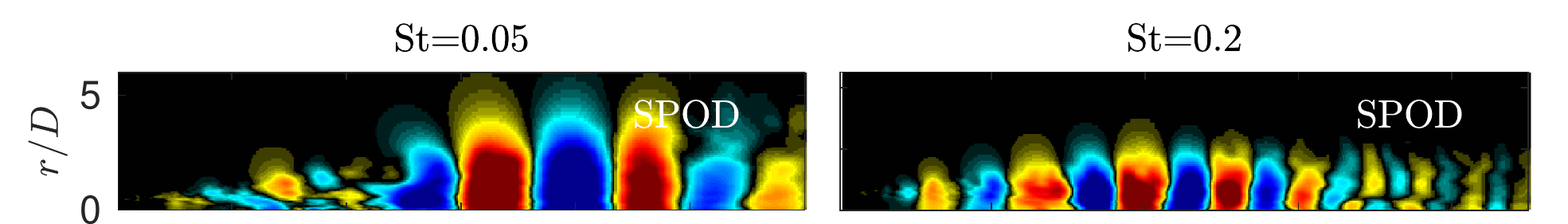}
\vspace{0cm}
\hrulefill
\vspace{0.25cm}
\includegraphics[width=1\textwidth,trim={0cm 0cm 0cm 3cm},clip]{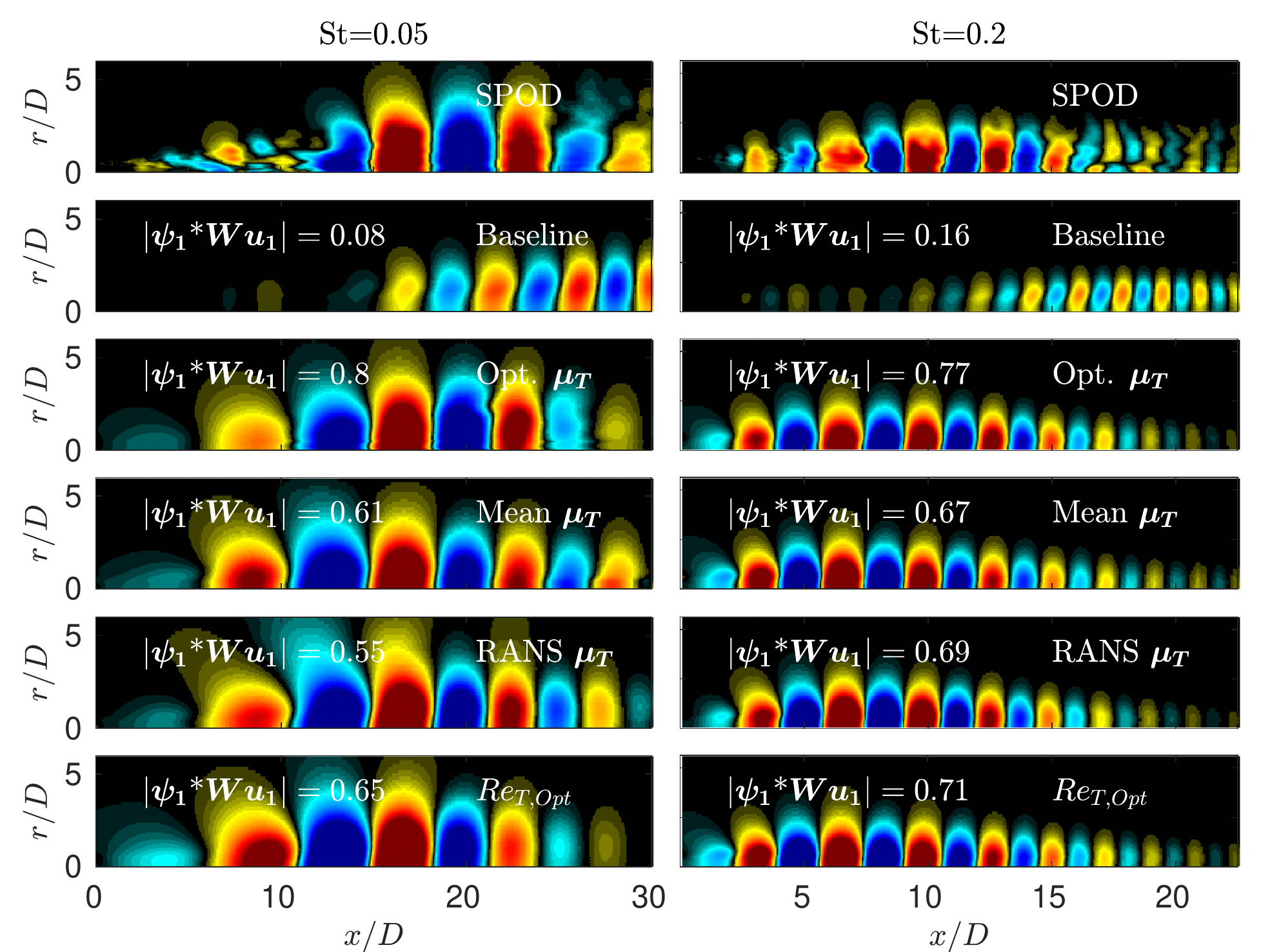}
\caption{Real component of the response pressure fluctuations (${\color{Red}\blacksquare}\!{\color{Black}\blacksquare}\!{\color{Blue}\blacksquare}$, $\pm 0.5||\bm{\psi}_1:p||_\infty$) for $St=0.05$ and $St=0.2$ in the left and right columns, respectively. Row 1 presents the dominant SPOD mode for which the optimization seeks to match. The following rows present results for the baseline, optimal eddy-viscosity field, mean-flow consistent model, RANS eddy-viscosity model, and the optimal turbulent Reynolds number.}
\label{fig:St005_02}
\end{figure}

\begin{figure}
\centering
\vspace{0.5cm}
\includegraphics[width=1\textwidth,trim={0cm 12.7cm 0cm 0cm},clip]{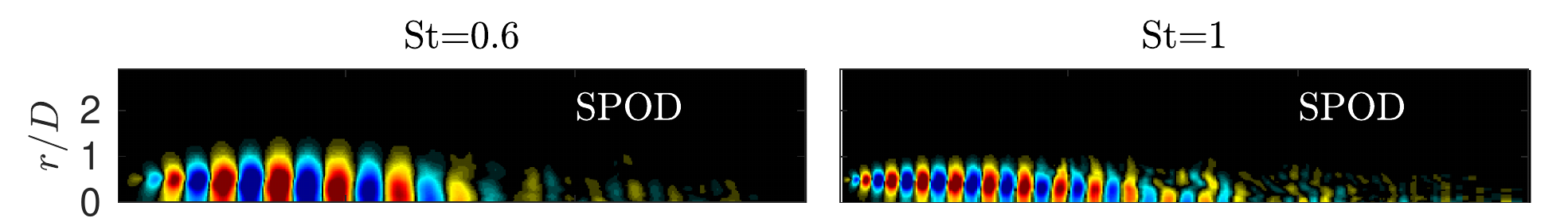}
\vspace{0cm}
\hrulefill
\vspace{0.25cm}
\includegraphics[width=1\textwidth,trim={0cm 0cm 0cm 3cm},clip]{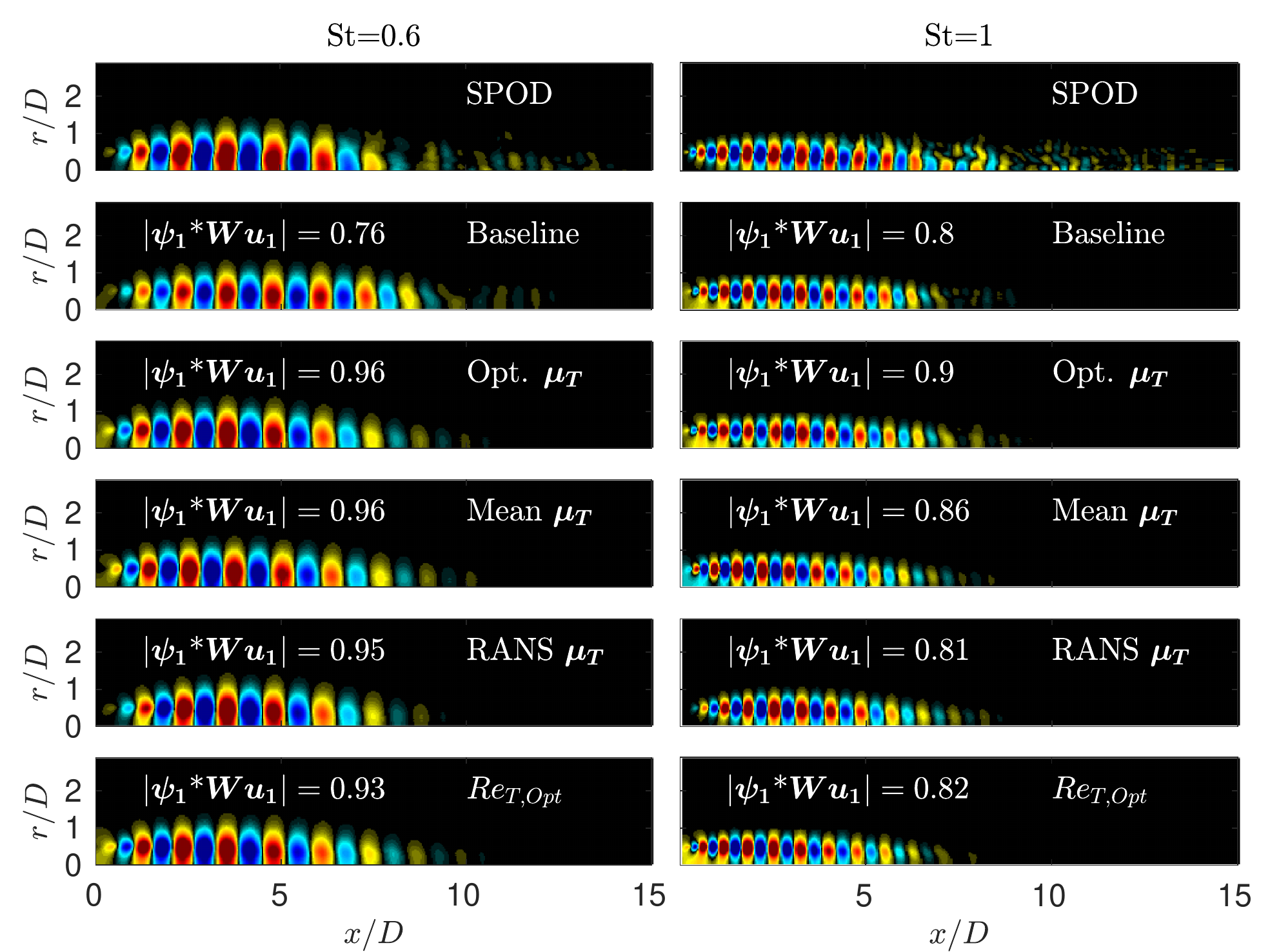}
\caption{Real component of the response pressure fluctuations for $St=0.6$ and $St=1$ in the left and right columns, respectively. Rows present the equivalent methods as described in figure \ref{fig:St005_02}.}
\label{fig:St06_1}
\end{figure}

Starting with the lowest frequency, $St=0.05$, we now investigate the mode shapes associated with the improved resolvent alignments achieved with the optimized eddy-viscosity models. Figure \ref{fig:St005} displays the real part of the fluctuating field for all state variables for the dominant SPOD and resolvent modes, comparing resolvent results using both the optimal eddy-viscosity field and the baseline case with constant $Re_T = 3 \times 10^4$. It is immediately apparent that the optimal eddy-viscosity resolvent mode can closely match the observed mode shapes from SPOD for all variables (including the correct distribution of energy), while the baseline resolvent mode bears little resemblance to the SPOD modes for any of the variables.   

Despite the increased alignment, there remains an obvious mismatch in $u_\theta’$ between the SPOD and resolvent modes, highlighting a statistical limitation to our approach. For the axisymmetric wavenumber, $m=0$, perturbations in the azimuthal velocity must be zero. Both resolvent models meet this constraint, however, the SPOD mode does not. One should then view the nonzero component in the SPOD mode as a statistical error. Compared to the streamwise velocity, $u_\theta’$ is about 5 times smaller in magnitude, and lacks the coherent wavepacket structure of the other variables. The corresponding $u_\theta’$ contribution in the projection coefficient $|\bm{\psi}_1^* \bm{W} \bm{\psi}_1|$ is $\approx 0.08$, bounding the physical maximum of the optimization to $|\bm{\psi}_1^* \bm{W} \bm{u}_1| \leq 0.92$ without considering additional error in the other variables. We link these statistical errors to the weak low-rank behavior with this frequency, where there is little eigenvalue separation between the dominant and subdominant modes \citep{schmidt2018spectral}. We may then view the projection-coefficient value of 0.08 as a kind of error bar on the alignments produced by the optimal eddy-viscosity field, as it is attempting to align to a mode shape that is (at this frequency) in error by as much as about 10\%.

The pressure field, a quantity of particular interest for jet noise, provides a relatively simple representative mode shape for each case.  We proceed by visualizing only the fluctuating pressure component for the rest of the study, however, the projection coefficients, $|\bm{\psi}_1^* \bm{W} \bm{u}_1|$, account for the full state results. Further, for all response pressure modes presented, we see similar trends and improvements in all flow variables similar to figure \ref{fig:St005}.

Figure \ref{fig:St005_02} shows the pressure modes at two low frequencies, $St=0.05$ and 0.2, and compares the results for all considered eddy-viscosity models. The top row shows the dominant SPOD mode from the LES, the second row gives the dominant resolvent mode for the baseline case, and the remaining rows provide the four optimized models.  At low frequencies, the baseline resolvent analysis cannot capture the observed mode shapes, while the optimized eddy-viscosity models have much better alignment with SPOD. The EVRA models increase the projection coefficients by as much as 10-fold and display a wavepacket structure consistent with the SPOD mode.  Orr-type modes dominate the low-frequency (i.e. $St < 0.3$) baseline resolvent spectrum \citep{schmidt2018spectral}, and we see that the eddy viscosity attenuates these modes in favor of a KH-like response that peaks further upstream, consistent with the observed SPOD modes.

Proceeding to higher frequencies, figure \ref{fig:St06_1} displays the dominant fluctuating pressure modes for SPOD and the five EVRA models for $St = 0.6$ and 1. The baseline projection coefficients are already high for these frequencies, but are further increased with the eddy-viscosity models, reaching 96\% for the optimal eddy viscosity.  Here the differences in the mode shapes are subtle, with the streamwise extent of the modes shortening from the baseline case to better match the SPOD at both frequencies.  At these higher frequencies, the jet response is a clear, low-rank KH wavepacket (a modal, inviscid stability mechanism) and it is thus unsurprising that the results are relatively insensitive to the precise eddy-viscosity model. However, the improved alignment is a product of the non-zero eddy-viscosity field, showing that a turbulence model is still important.

For $St=1$, the optimized projection coefficient is falling compared to the $St=0.6$ case. This is due to the emergence of Orr-type modes with similar energy as the KH modes. When performing SPOD in limited domains near the nozzle exit, the modal, low-rank KH response continues to dominate at much higher frequencies in the near nozzle region \citep{sasaki2017high}, but when considering the global response, the KH response becomes inferior, in energy, to the Orr response, which peaks further downstream.

\section{Analysis of the optimized eddy-viscosity fields} \label{sec:Opt_params}

\begin{figure}
\hspace{0.25cm} (a) \hspace{0.8cm}  $Re_{T,Opt.}^{-1}$ \hspace{0.4cm} (b) \hspace{0.4cm} Opt.  $\norm{\bm{\mu}_T}_{\infty}$ \hspace{0.25cm} (c) \hspace{0.7cm} Mean $c$ \hspace{0.675cm} (d) \hspace{0.7cm}  RANS $c$ \\
 \includegraphics[width=1\textwidth,trim={0cm 0cm 0cm 0.5cm},clip]{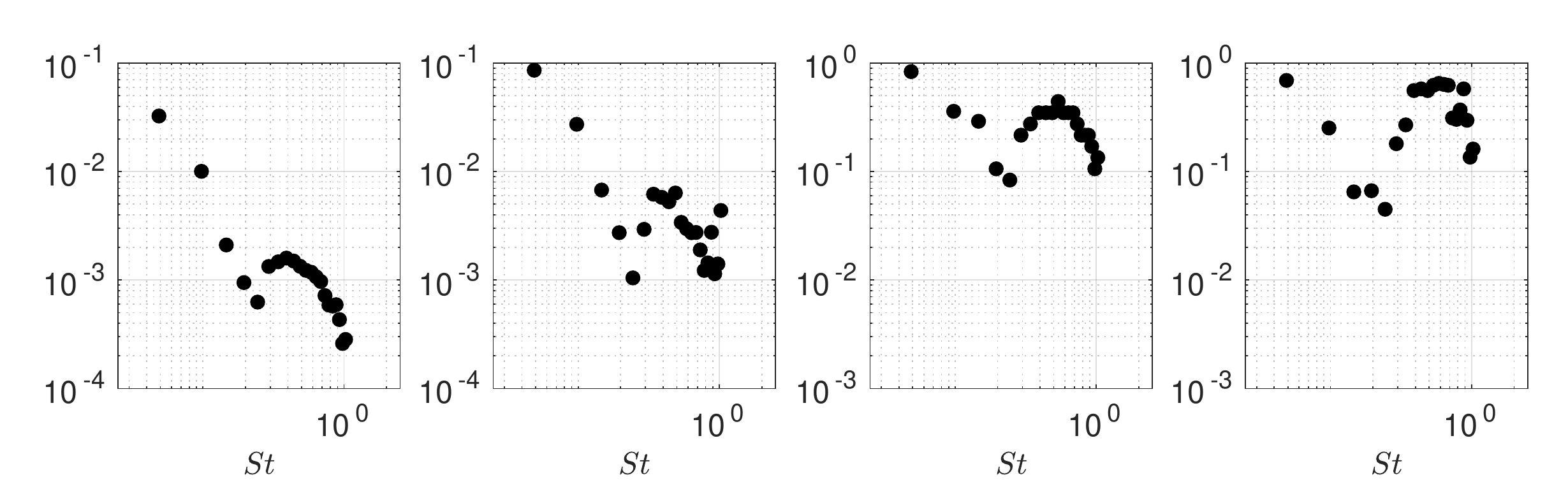} 
 \caption{The optimal parameters across $St  \in[0.05,1]$ for (a) the optimal constant field $1/Re_T$, (b) optimal eddy-viscosity field model, (c) the mean-flow consistent model, and (d) the optimal RANS model. The optimal eddy-viscosity field parameter shown is the maximum value of the field at each frequency, $\norm{\bm{\mu}_T}_{\infty}$, while the latter two models present the optimal coefficient $c$. The associated alignments for each model/parameter are shown in figure \ref{fig:alignments}. }
 \label{fig:Optimal_Params}
\end{figure}

The previous section shows that the EVRA approach results in substantial alignment of the dominant resolvent and SPOD modes. In this section, we examine the optimal parameters associated with the eddy-viscosity fields to investigate {\it how} the eddy viscosity improved the alignment and to identify potential universalities in modeling coefficients.

\subsection{Structure of the eddy-viscosity fields} \label{sec:mult_structure}

For the constant eddy viscosity, RANS-based, and mean-flow consistent eddy-viscosity fields, the optimization is over a single value, and we plot the optimal values as a function of frequency (still for $m=0$) in figure \ref{fig:Optimal_Params}~(a,c,d) and the maximum value of the optimal field in \ref{fig:Optimal_Params}~(b). We investigated several other metrics for the optimal field and each metric provided similar trends and therefore, we chose $\norm{\bm{\mu}_T}_{\infty}$, as it gave the most intuitive comparison against the other scalar quantities. For all models, the frequency dependence of the values are similar, with three regions of interest: $St  \in[0.05,0.3]$, $St =  \in[0.3,0.8]$, and $St  \in[0.8,1]$. 

In the low frequency region, the baseline jet response comprises of spatially extensive Orr-type modes that have a strong Reynolds number dependence, requiring a relatively larger eddy viscosity to damp them. For $St=0.05$ the ratio of the molecular Reynolds number to the optimal effective Reynolds number is $\mu_j/\bm{\mu}_T \approx 13,500$, a four order-of-magnitude difference when compared to the molecular viscosity. 

In the moderate frequency regime, where the baseline spectrum transitions from the broadband, viscous Orr mechanism to the low-rank, inviscid KH mechanism, eddy viscosity becomes less important, and we expect (confirming below, in \S~\ref{sec:Sensitivity}) insensitivity to the overall value based on the relatively favorable alignment achieved in the baseline case. As frequency increases, the responses transition back to a mix of KH and Orr-type waves, with a progression towards broadband, viscous Orr modes at higher frequency. 

At these higher frequencies, we see that the low-frequency dependence on inverse effective Reynolds number resumes, similar to low the frequencies. Interestingly, this trend shows that at higher frequencies $Re_T \rightarrow Re_j $ such that the effect of eddy viscosity ``turns-off’’ as frequency increases and the associated wavepacket wavelength becomes small (i.e. approaching finer-scale turbulence), as expected on physical grounds.

\begin{figure}
\centering
\vspace{0.5cm}
\includegraphics[width=1\textwidth,trim={0cm 0cm 0cm 0cm},clip]{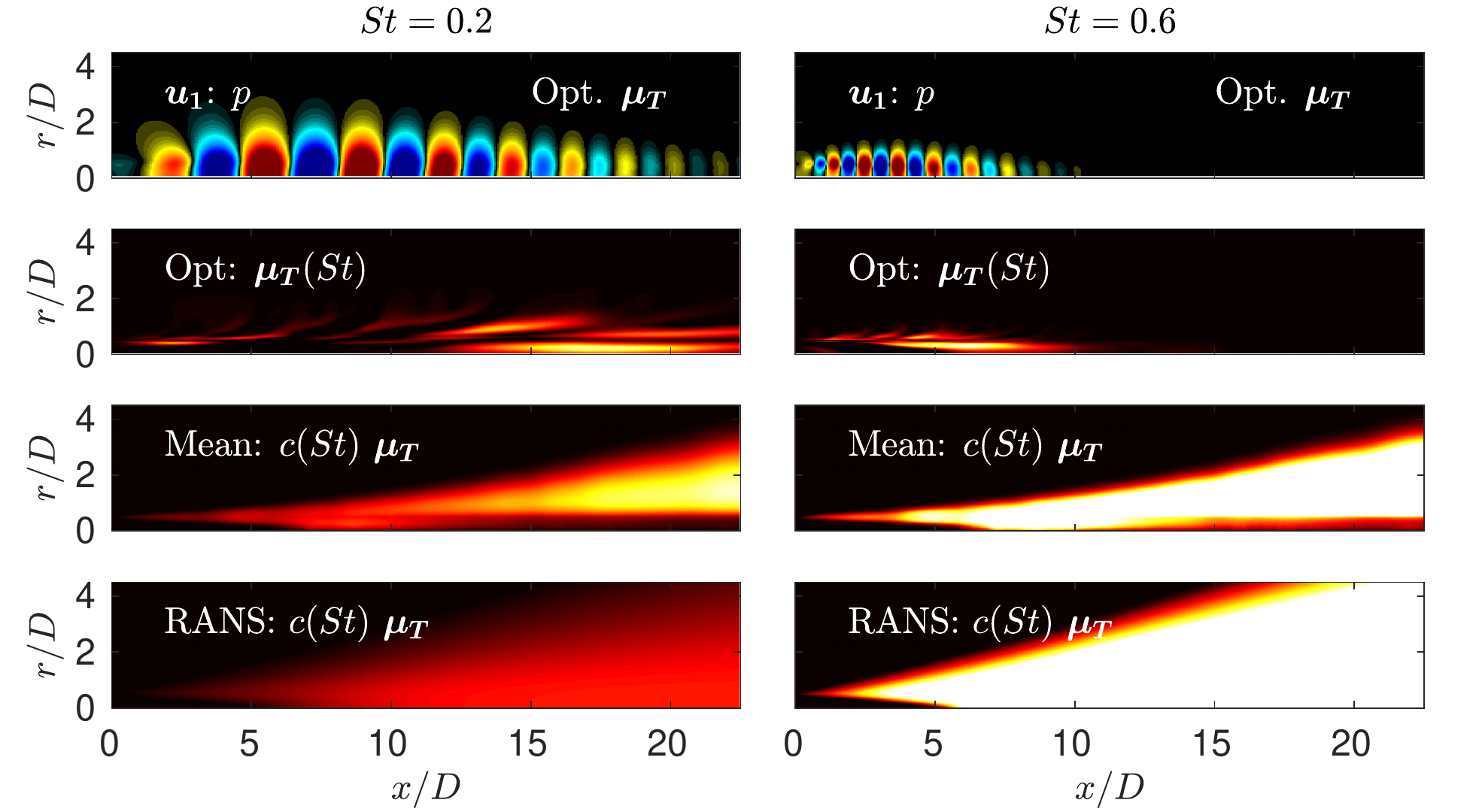}
\caption{Comparisons of the optimal eddy-viscosity fields (i.e. full-field optimal, mean-flow consistent, and RANS) and the associated dominant resolvent mode found via the optimization for $St = 0.2$ and $0.6$. Contours for all six eddy-viscosity fields are set from 0 to $3 \times 10^{-3}$.}
\label{fig:Opt_Visc_Fields}
\end{figure}

For the full-field eddy-viscosity optimization, we stress that its primary purpose is to determine what may be an upper bound for how well \textit{any} eddy-viscosity model could perform. Given that the alignments between the resolvent and SPOD modes were not significantly higher for the optimized scheme than for the modeled eddy-viscosity approaches (with optimal parameters), the detailed eddy-viscosity fields are of lesser importance. Still, some aspects of the physics, such as the spatial locations where Reynolds stresses become important for each frequency, are apparent in the optimized fields. Figure \ref{fig:Opt_Visc_Fields} presents the optimized fields for two selected Strouhal numbers, comparing them to both the RANS and mean-flow consistent eddy-viscosity fields scaled by their optimal coefficient $c$ at each frequency. In addition, the dominant resolvent mode, computed with the displayed optimal-eddy-viscosity field, is shown for comparison with the eddy-viscosity fields. The contour for the eddy-viscosity fields are set from 0 to the maximum value of the $St = 0.6$ optimal eddy-viscosity field. 

Overall, both frequencies present optimal eddy-viscosity fields that are complex, unsurprising given the ability of the optimization to choose any eddy-viscosity field, constrained only by the structure of the equations and positivity. The optimal eddy-viscosity fields pinpoint the locations where linear structures break down (i.e. where nonlinearities/Reynolds stresses become important) and inform what features an eddy-viscosity model must include. In both cases, the optimization removes viscosity from the potential core (i.e. the interior region of the jet relative to the critical layer), when compared to the initial guess, while increasing the turbulent viscosity just outside of the critical layer. The increase in eddy-viscosity is most often observed just downstream of the peak amplitude of the wavepacket, coinciding with each wavepacket’s decay downstream. 

Although not entirely clear from figure \ref{fig:Opt_Visc_Fields}, these findings are reasonably consistent with each of the modeled eddy-viscosity fields when restricting the view to the region where the resolvent/SPOD mode has significant amplitude. We can see that both the RANS and mean-flow consistent eddy-viscosity fields present similar features as the optimal field, explaining the ability of each model to achieve nearly optimal results. We will show in the following section how such features also explain the ability of the RANS and mean-flow consistent models to predict the subdominant modes, which require further turbulence modeling downstream.

\section{Alignment of subdominant modes}\label{sec:subdominant}

\begin{figure}
\centering
\vspace{0.5cm}
\includegraphics[width=1\textwidth,trim={0cm 9.15cm 0cm 0cm},clip]{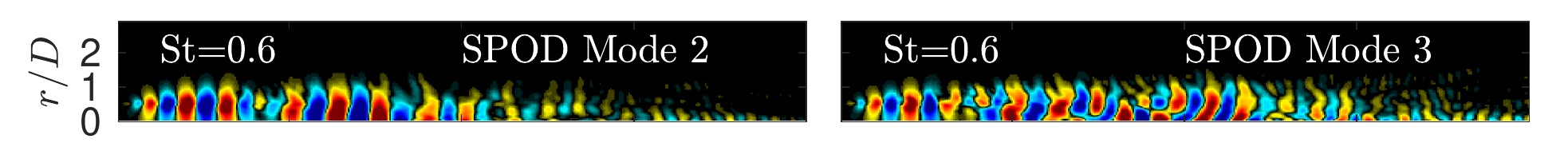}
\vspace{0cm}
\hrulefill
\vspace{0.25cm}
\includegraphics[width=1\textwidth,trim={0cm 0cm 0cm 1.75cm},clip]{figs/fig10.pdf}
\caption{Subdominant modes 2 and 3 at $St = 0.6, m=0$ in the left and right columns respectively for SPOD, baseline, and all EVRA models.}
\label{fig:SubOpts1}
\end{figure}

Although the optimization presented only aligns the dominant SPOD and resolvent modes, subdominant modes are also of interest, particularly as they are necessary to reconstruct flow statistics in the near field and are relevant for modeling coherence decay associated with the `jittering of wavepackets' to produce sound \citep{cavalieri2011jittering}. In this section we seek to answer two questions, whether alignment with only the dominant mode substantially alters the alignment of the subdominant modes and the effect of expanding the optimization to subdominant modes. We first assess the former case using the optimal parameters for each method. We show the computed subdominant modes in figure \ref{fig:SubOpts1} for modes 2 and 3 for the $St = 0.6$, $m=0$ frequency-wavenumber pair.

\begin{figure}
\centering
\vspace{0.5cm}
\includegraphics[width=1\textwidth,trim={0cm 0cm 0cm 0cm},clip]{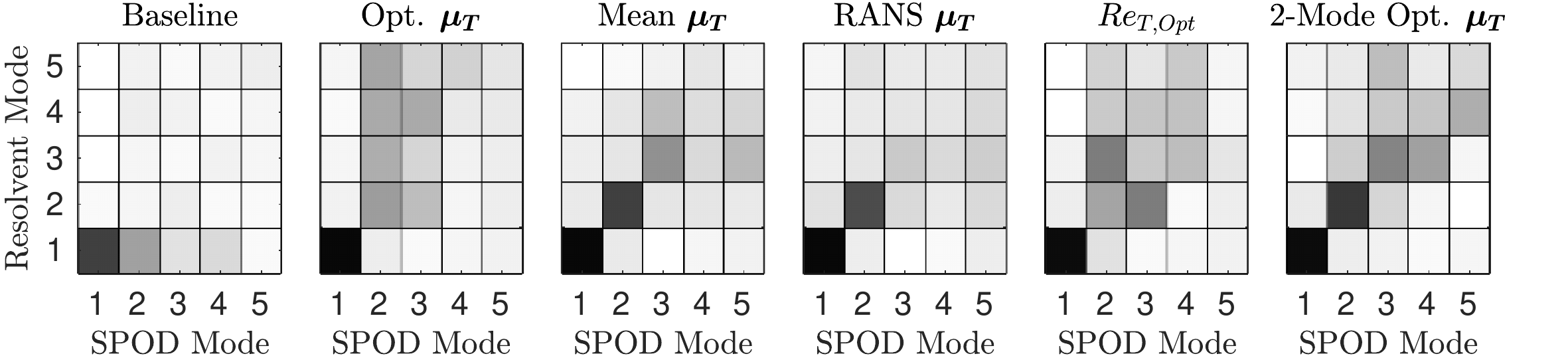}
\caption{Projections of the first five SPOD modes into the first five resolvent modes computed for all EVRA models at $St=0.6, m=0$, including the 2-mode optimization shown in figure \ref{fig:SubDom2}.}
\label{fig:Projections}
\end{figure}

Comparing the second mode to the baseline case ($Re_T = 3 \times 10^4$), we find that all EVRA models give significantly improved alignments, reaching $\approx$70\% for the RANS and mean-flow consistent models. Both the RANS and mean-flow consistent models are superior to the optimal eddy-viscosity field, which is only fitted to align the dominant mode. The RANS and mean-field models are also superior for the third, fourth and fifth modes (the latter two not shown for brevity), but with an alignment that falls off with increasing mode number. 

To observe how well the optimization of the first SPOD mode models the forcing statistics (i.e. diagonalizes the forcing CSD $\bm{S}_{\bm{ff}}$), we compare projections of the first five SPOD modes with the first five modes from each eddy-viscosity method (including a 2-mode optimization described next) in figure \ref{fig:Projections}. The plots show that the EVRA models, in particular the RANS and mean-flow consistent models, are superior at diagonalizing the CSD when compared to the baseline case.

Although the optimal eddy-viscosity field, aligned only with the dominant SPOD mode, shows improvements in the subdominant modes, we can extend the optimization to align an arbitrary number of subdominant modes and achieve alignment superior to any eddy-viscosity model. However, convergence issues with increasing SPOD mode number suggest that optimizing for many modes (e.g. $n>5$) would have marginal returns. For this study, we present only the optimization of both the first and second modes at $St = 0.6$, $m=0$ to show the generality of the optimization framework and the physical implications of the associated eddy-viscosity field for the subdominant modes.

\begin{figure}
\centering
\includegraphics[width=1\textwidth,trim={0cm 0cm 0cm 1cm},clip]{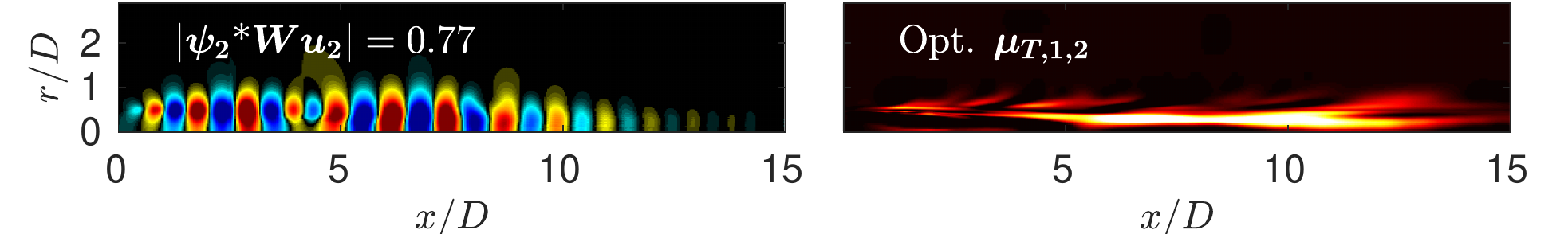}
\caption{The second subdominant mode at $St=0.6$ and the associated eddy-viscosity field that provides the optimal alignment for both modes. The contour for the eddy-viscosity field is set to the same value as those shown in figure \ref{fig:Opt_Visc_Fields} from 0 to $3\times10^{-3}$. }
\label{fig:SubDom2}
\end{figure}

Figure \ref{fig:SubDom2} presents the aligned resolvent mode via the optimization and the associated eddy-viscosity field for the first subdominant mode. By including the second SPOD mode, the optimization can achieve an alignment of 77\%, superior to any of the other eddy-viscosity models, without altering the alignment of the dominant mode, 96\%. We also observe that the remaining subdominant modes also increase in their projections, as shown in figure \ref{fig:Projections}. This observation is likely linked to the difference in mechanisms of the dominant and subdominant modes at $St = 0.6$, $m=0$.  The dominant mode is KH-type, while the subdominant modes are of Orr-type. By aligning just the first Orr-type mode, we observe improved alignments for the entire family of Orr modes, conversely, alignment of only the KH mode does not substantially improve Orr modes.

The increase in alignment results from additional eddy-viscosity located downstream of the 1-mode, KH-type field, $\bm{\mu}_{T,1}$, shown in figure \ref{fig:Opt_Visc_Fields}. The second mode imposes a need for further eddy-viscosity acting further downstream and towards the centerline, as representative of the Orr-mechanism at $m=0$ for turbulent jets \citep{pickering2020lift}. We find that this additional downstream eddy-viscosity, present in both the RANS and mean-flow consistent models, is responsible for the increased subdominant mode alignment. Considering the simpler RANS (and mean-flow) model also shows similar downstream structure, we investigate its merit for a predictive model in the next section.

\section{Towards a predictive EVRA model for turbulent jets}\label{sec:complete}

Through the previous sections, we have shown that both the RANS and mean-flow consistent eddy-viscosity models perform well across Strouhal numbers from 0.05 to 1 at $m=0$, provided the overall constant associated with their application to the disturbance fields is optimal (at each frequency and azimuthal mode number).  In this section, we consider the sensitivity of the results regarding the choice of a frequency (and wavenumber) independent constant, and show that over a range of frequencies and azimuthal mode numbers, alignments are relatively insensitive to the choice of a constant, such that a single, universal value may be acceptable.  While both RANS and mean-flow consistent models both performed well with optimal coefficients, we focus only on the RANS $k-\epsilon$ model, as it is better regarded as universal across a range of flows. We then apply EVRA-RANS to the $M_j=0.4$ jet using a single constant to six azimuthal wavenumbers, $m = 0-5$, and find substantially improved predictions when compared to the baseline. We also find similar observations when using the same EVRA-RANS model for both the transonic and supersonic jets. Finally, we present the effect of the eddy viscosity on the resolvent spectra.

\subsection{Frequency and azimuthal mode sensitivity} \label{sec:Sensitivity}

\begin{figure}
\centering
  \includegraphics[width=0.5\textwidth,trim={0cm 0cm 0cm 0cm},clip]{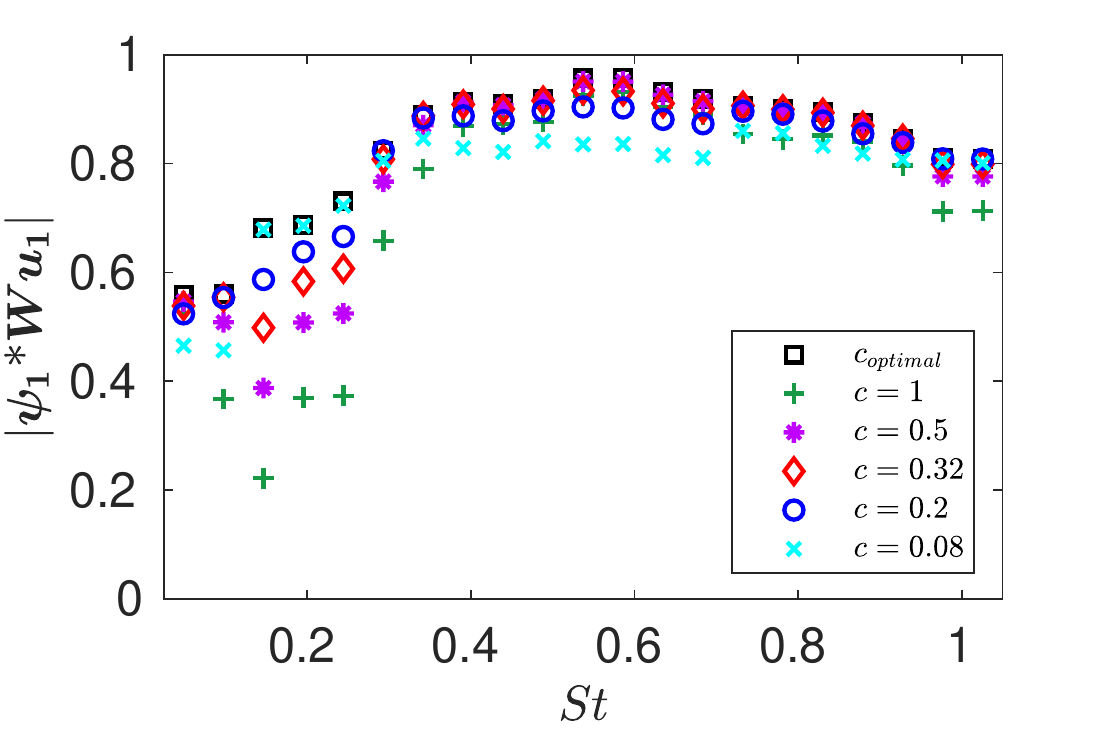}
\caption{Alignments across all Strouhal numbers for the RANS eddy-viscosity model coefficients compared with the optimal RANS coefficient at each frequency. The RANS coefficients are $c = [1, 0.5, 0.32, 0.2, 0.08]$.}
\label{fig:RANS_alignments}
\end{figure}

The optimal RANS coefficients (figure \ref{fig:Optimal_Params}) ranged from $c = 0.7-0.004$, with a relatively constant region,  $c=0.5$, for moderate frequencies and, considering the fully optimized eddy-viscosity field produced only marginally improved alignments for most cases, the results may not be sensitive to the precise constant. We test this hypothesis for the RANS model across a range of frequencies with proposed ``universal’’ values of constant $c = [1, 0.5, 0.32, 0.2, 0.08]$.  We plot the resulting alignments versus frequency in figure \ref{fig:RANS_alignments}. With little compromise, compared to the optimal constant for each frequency, a single constant of $c=0.2$ provides significant alignment across all frequencies up to $St=1$.  Although not shown for brevity, we found similar observations using $c=0.08-1$ for all three Mach numbers and six azimuthal wavenumbers. In these cases, not only did $c=0.2$ give the best overall alignment, but the alignments were comparably insensitive to the value of $c$ chosen over this range.

\begin{figure}
\hspace{0.35cm} (a) \hspace{1.35cm} RANS $c=0.2$ \hspace{2.55cm} (b) \hspace{1.75cm} Baseline \\
 \begin{minipage}{0.49\textwidth}
 \includegraphics[width=1\textwidth,trim={0cm 0cm 0cm 0 cm},clip]{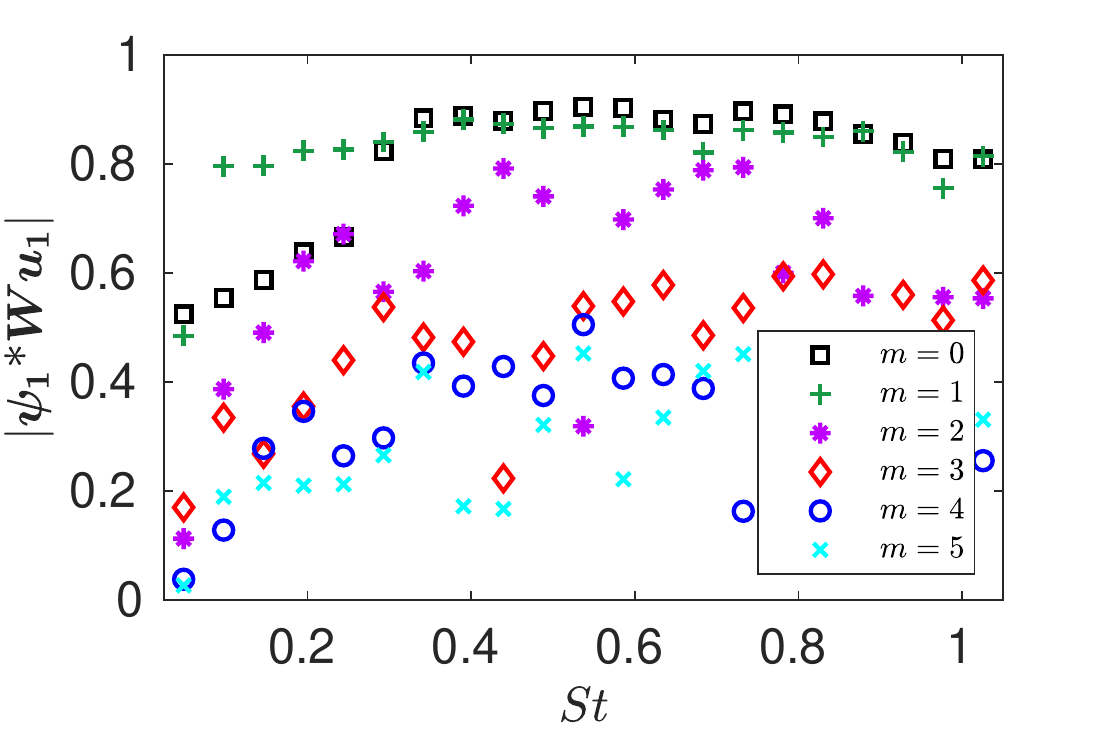}
    \end{minipage}
 \begin{minipage}{0.49\textwidth}
  \includegraphics[width=1\textwidth,trim={0cm 0cm 0cm 0cm},clip]{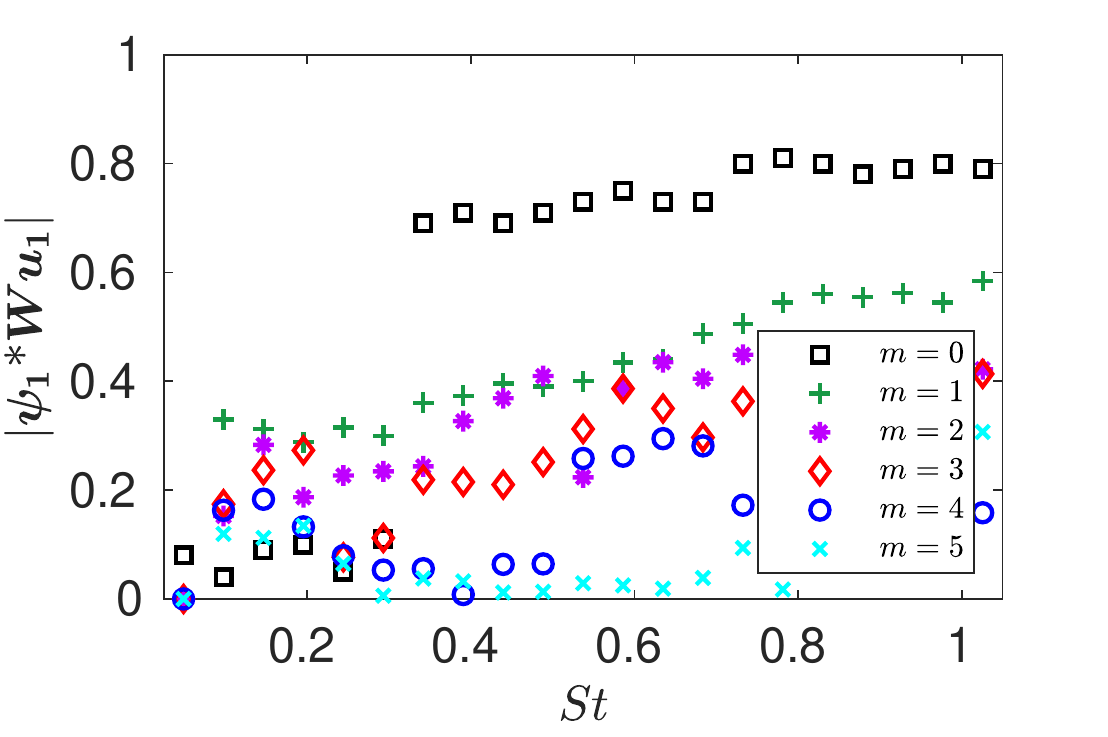}
\end{minipage}
\caption{Alignments for frequencies, $St  \in[0.05,1] $, and azimuthal wavenumbers, $m = 0-5 $, for the (a) RANS eddy-viscosity model using $c = 0.2$ and the (b) baseline, constant eddy-viscosity case (i.e. $Re_T = 3 \times 10^4$). }
\label{fig:RANS_m_alignments}
\end{figure}

For nonzero azimuthal modes, figure \ref{fig:RANS_m_alignments} presents the alignment of the EVRA-RANS model with SPOD using $c=0.2$ and the baseline case for $m = 0-5$. The EVRA-RANS model substantially increases the alignments for all nonzero wavenumbers. The results for $m=1$ are particularly encouraging, with a uniform, $80$\% alignment across all frequencies. Azimuthal modes greater than 1 result in poorer alignment, albeit much improved compared to the baseline case, especially when $m > 2$. 

Expanding to nonzero azimuthal wavenumbers, the eddy-viscosity field also affects a third mechanism observed in the global SPOD spectrum (as $St \rightarrow 0$), the lift-up mechanism \citep{pickering2020lift}. Similar to the Orr mechanism, the lift-up mechanism arises from triadic nonlinear interactions in the flow \citep{hamilton1995regeneration, sharma2013coherent,de2017streak,cho2018scale}, identifying it as a likely benefactor to an EVRA approach. Figure \ref{fig:RANS_m_alignments} supports this claim, showing significant improvements at low-frequencies for nonzero wavenumbers. These observations also agree with \cite{pickering2020lift}, who showed that resolvent modes related to streaks required an eddy-viscosity model (using the TKE model reported by \cite{pickering2019eddy} with $c=0.0065$). They also observed that, in turbulent jets, the spatial extent of resolvent modes increase as frequency decreases and that without an eddy-viscosity, modes extend indefinitely downstream for $St=0$. This is analogous to theory surrounding streaks where the lift-up mechanism presents a rapid spatial growth of streamwise streaks until viscous dissipation becomes dominant and the structures decay \citep{hultgren1981algebraic}.  Considering the significant improvements between alignments for low-frequency and nonzero wavenumbers, we find the lift-up mechanism to also be sensitive to an eddy-viscosity model.

\subsection{Transonic and supersonic turbulent jets} \label{sec:trans_super}

\begin{figure}
\hspace{0.35cm} (a) \hspace{1.75cm} $M_j = 0.9$ \hspace{2.85cm} (b) \hspace{1.75cm} $M_j = 1.5$\\
	\begin{minipage}{0.49\textwidth}
	\includegraphics[width=1\textwidth,trim={0cm 0cm 0cm 0 cm},clip]{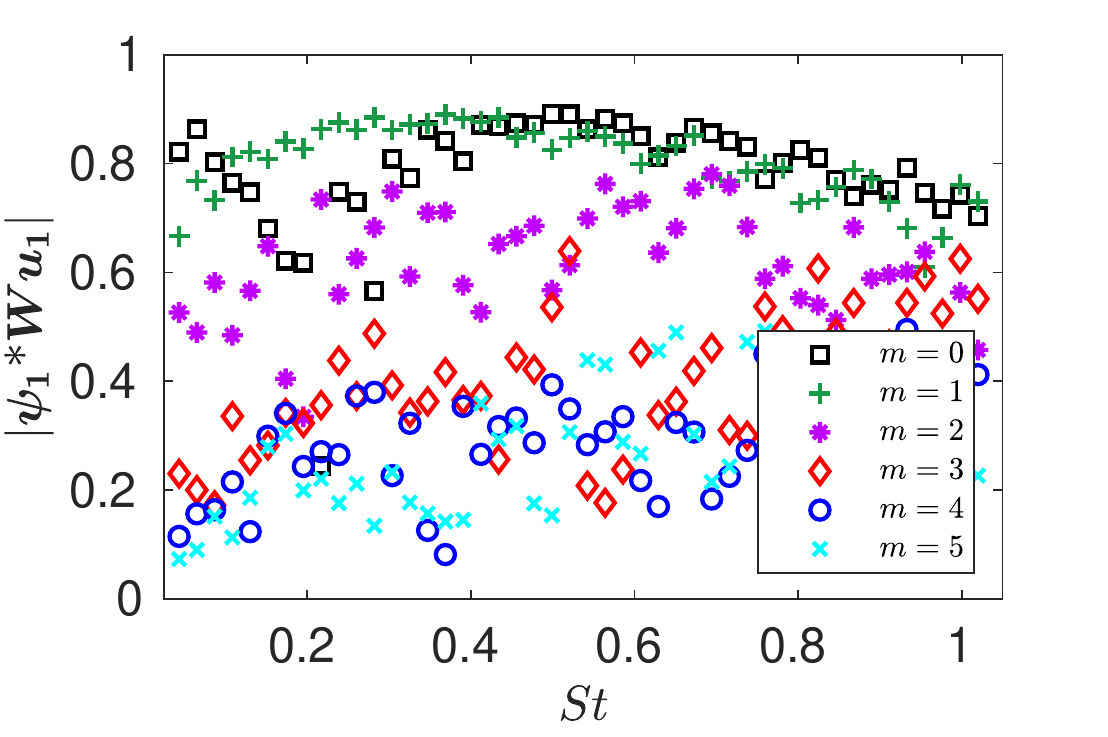}
    \end{minipage}
	\begin{minipage}{0.49\textwidth}
		\includegraphics[width=1\textwidth,trim={0cm 0cm 0cm 0cm},clip]{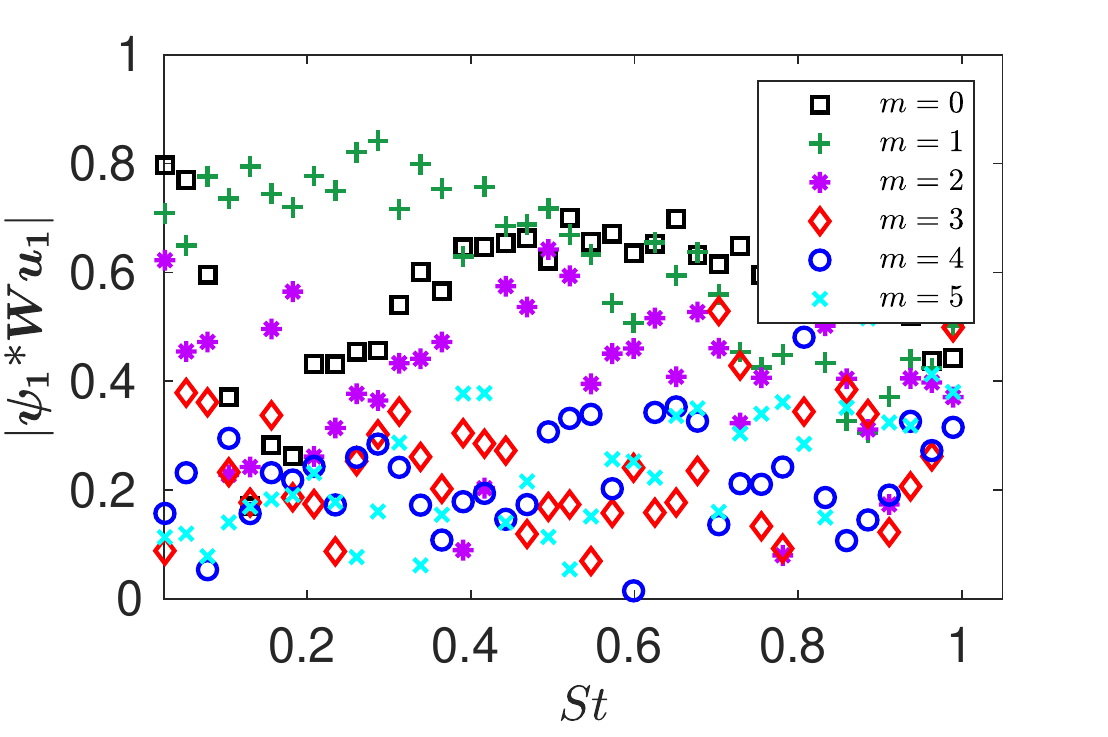}
\end{minipage}
\caption{Alignments using the RANS eddy-viscosity model with coefficient $c = 0.2$ across Strouhal numbers $St \in[0.05,1]$ and azimuthal wavenumbers $m =0-5$ for the (a) $M_j = 0.9$  and (b) $1.5$  jets.}
\label{fig:RANS_m_alignments_M09_M15}
\end{figure}

We now generalize the RANS-EVRA model performance for both $M_j = 0.9$ and 1.5 turbulent jets. Figure \ref{fig:RANS_m_alignments_M09_M15} provides the alignments across frequencies and azimuthal wavenumbers for each. The transonic jet gives substantial agreement for $m=0$ and $m=1$ at about 80\% for much of the frequency range, while $m=2$ gives alignments of 60\%, on average. For the supersonic jet, the agreement is not as favorable, however, much improved from the $Re_T = 3 \times 10^4$ alignments (not shown here for brevity). 

\begin{figure}
\centering $M_j = 0.4$\\
\centering \hspace{0.0225\textwidth} $m=0$ \hspace{0.0775\textwidth}$m=1$ \hspace{0.0775\textwidth}$m=2$\hspace{0.0775\textwidth}$m=3$ \hspace{0.0775\textwidth}$m=4$ \hspace{0.0775\textwidth} $m=5$\\
\includegraphics[width=1\textwidth,trim={0cm 1.3cm 0cm 0.75cm},clip]{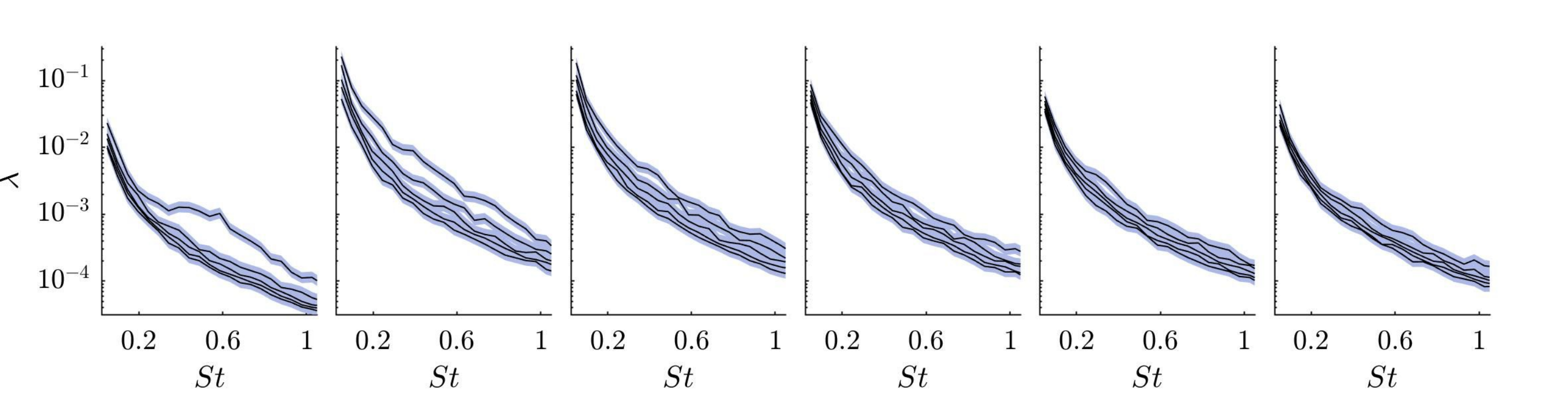}\\
$M_j = 0.9$\\
\includegraphics[width=1\textwidth,trim={0cm 1.3cm 0cm 0.5cm},clip]{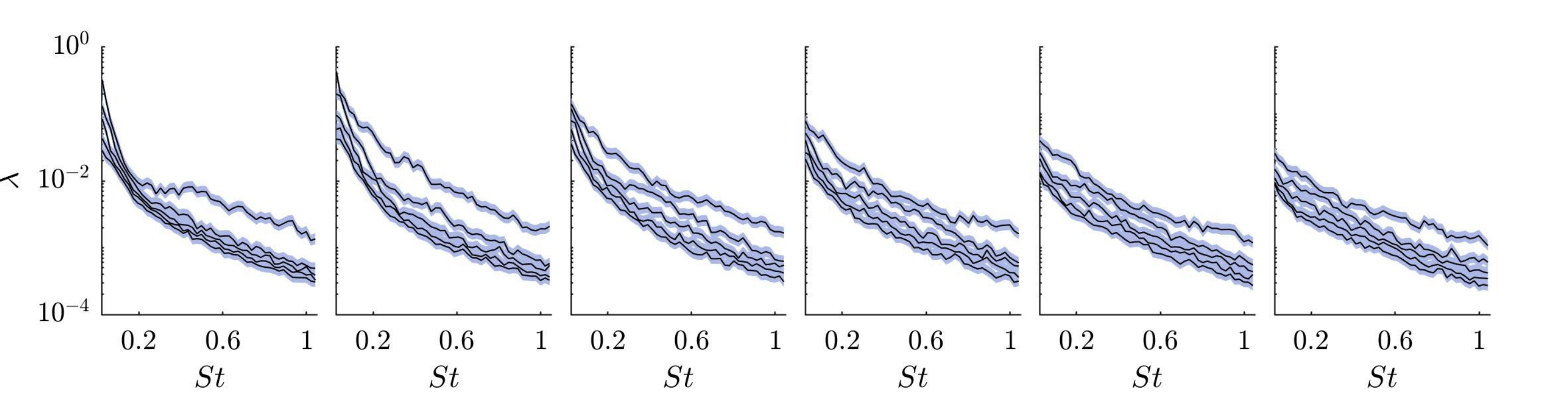}\\
$M_j = 1.5$\\
\includegraphics[width=1\textwidth,trim={0cm 0cm 0cm 0.5cm},clip]{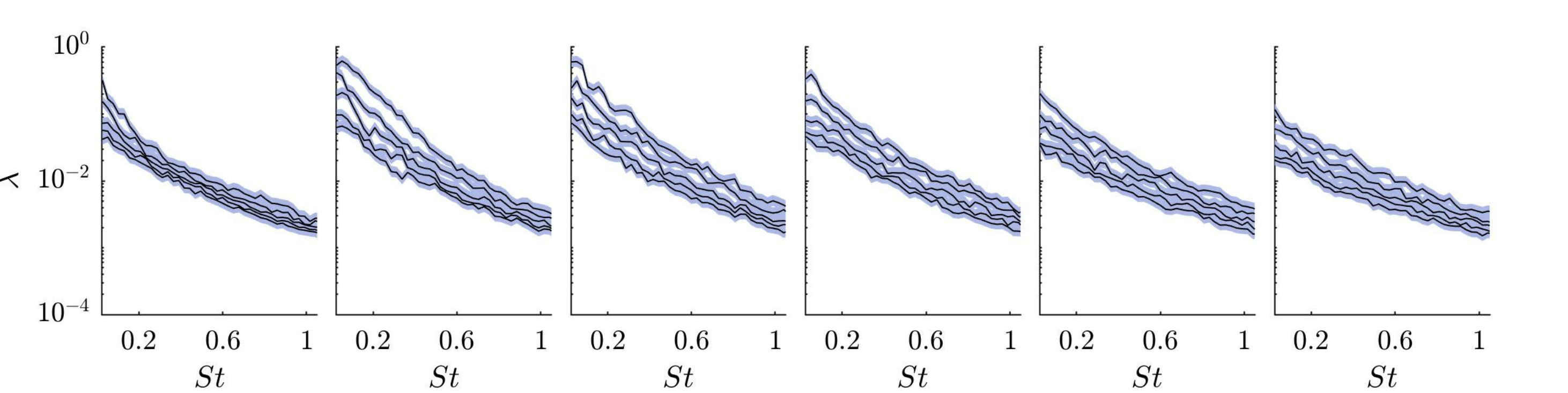} 
\caption{Spectra, and their associated 95 \% confidence interval in light blue, of the first 5 SPOD modes for azimuthal wavenumbers $m=0-5$ from left to right and the subsonic, transonic, and supersonic jets from top to bottom, respectively.}
\label{fig:SPOD_Spectra}
\end{figure}

The RANS eddy-viscosity model increases many of the alignments, however, poor alignments remain, and these alignments appear to correspond to SPOD spectra without large energy separation. As shown earlier in figure \ref{fig:alignments}, EVRA and SPOD modes aligned best when there exists large eigenvalue separation between the first and second SPOD mode. We find similar behavior here for all cases. Figure \ref{fig:SPOD_Spectra} presents the SPOD spectra of the first 5 modes across all six azimuthal wavenumbers and three turbulent jets, with their associated 95 \% confidence intervals in light blue. A handful of the spectra show a clear separation between mode energies, such as those between the first and second mode for $M_j = 0.4$, $m=0$ and 1, for $M_j = 0.9$, $m=0$ and 1 (and higher frequencies for $m = 2-5$), and for $M_j=1.5$, $m=1$.  In each case where there is large eigenvalue separation, we find, from figures \ref{fig:RANS_m_alignments} and \ref{fig:RANS_m_alignments_M09_M15}, significantly greater agreement in projection coefficients between the resolvent and SPOD modes, while finding poor projections for cases without clear separation in eigenvalues. We also observe this for the subdominant modes investigated in the $M_j=0.4$, $m=0$ case in \S~\ref{sec:subdominant}.

These observations point to a limitation to our method when comparing EVRA modes with SPOD modes. For the SPOD modes without clear eigenvalue separation, the eigenvalues themselves fall within the uncertainty bands (i.e. 95 \% confidence interval) of the other modes. The eigenvectors corresponding to these eigenvalues are expected to have, at best, similar uncertainty levels.  Thus, without more data, it is not possible to attribute the lack of agreement to a failure of the EVRA ansatz. 

\subsection{Singular values}

\begin{figure}
\hspace{0.25cm} (a) \hspace{1.1cm} SPOD \hspace{1.6cm} (b) \hspace{1.1cm} Baseline \hspace{1.45cm} (c) \hspace{0.65cm} RANS $c=0.2$ \\
\includegraphics[width=\textwidth,trim={1cm 0cm 1.15cm 0cm},clip]{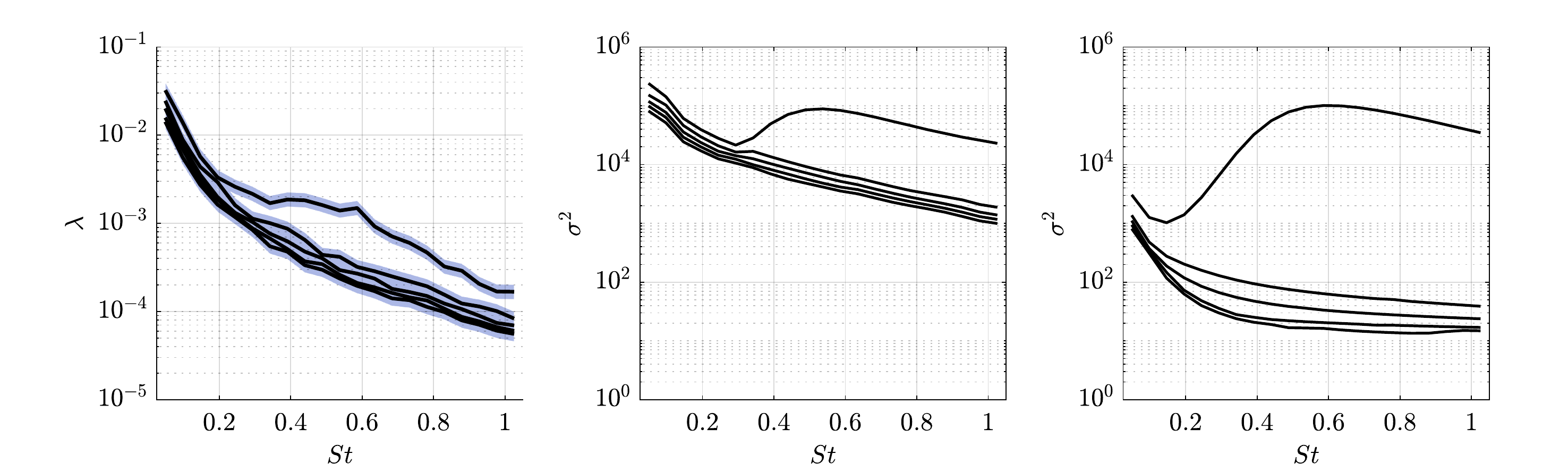}
\caption{Spectra of first five (a) SPOD, (b) baseline resolvent, and (c) the RANS eddy-viscosity model resolvent modes at $m=0$ for $St  \in[0.05,1]$.}
\label{fig:Spectrums}
\end{figure}

We return to the $m=0$, $St \in [0.05,1]$ case to assess the EVRA-RANS $c=0.2$ model’s effect on the singular values and compare them to the baseline case and the SPOD eigenvalues. Figure \ref{fig:Spectrums} provides the spectra of the first five modes for SPOD (accompanied by a shaded region providing the 95\% confidence interval of the eigenvalues), the baseline resolvent model, and the RANS-EVRA model (using $c=0.2$) for $m=0$. Comparing the resolvent spectra to the SPOD spectra, we immediately see that the separation between $\lambda$ (i.e. the ratio between $\lambda_n /\lambda_{n+1}$) and $\sigma^2$ of either resolvent models does not compare favorably. In fact, the RANS-EVRA spectra has increased its energetic separation when compared to the baseline case. 

This behavior may be linked to multiple (in this case two for $m=0$) distinct mechanisms represented in the flow, the KH and Orr-mechanisms. As detailed earlier, the inclusion of an eddy-viscosity model presents a substantial effect on the Orr modes significantly reducing the streamwise extent of each mode, while the KH modes are relatively unchanged. We observe an analogous effect here in figure \ref{fig:Spectrums} where the singular values related to the Orr mechanism decrease substantially, pulling away from the unaffected singular values of the KH mechanism, resulting in a much larger separation between singular values than is observed between the SPOD eigenvalues. This sensitivity of Orr modes to an eddy-viscosity was also observed in \cite{schmidt2018spectral} at $St=0.6, m=0$ when adjusting $Re_T$, finding that the squared singular values of the subdominant Orr modes scaled as $Re_T^{1.2}$. We observe the same effect using the RANS eddy-viscosity model, interestingly (and perhaps unsurprising given the preceding discussions), figure \ref{fig:Spectrums} (c) provides similar values as those reported by \cite{schmidt2018spectral} at $St = 0.6, m=0$ when using $Re_T = 10^3$.

Figure \ref{fig:Spectrums} (a) and (c) also show that the forcing amplitudes, $\lambda_\beta$, are not uniform in turbulent jets, contrary to a customary assumption used in resolvent analysis where $\bm{\Lambda_{\bm{\beta}}} = \alpha \bm{I}$, with $\alpha$ as an arbitrary constant \citep{morra2019relevance,lesshafft2019resolvent,hwang2020attached}. Focusing on only the first and second resolvent and SPOD modes for $St=0.6$ and $m=0$, where mode alignments are 95\% and 69\%, respectively (the optimal-field case increases the latter value to 77\% without appreciably changing the singular value), we may assume that the diagonal components of the forcing, $\lambda_{\bm{\beta},1}$ and $\lambda_{\bm{\beta},2}$, account for nearly all the energetic contributions by these two modes. As shown by equation \eqref{eqn:betaI}, this assumption allows for a one-to-one comparison between the first two SPOD eigenvalues, resolvent singular values, and forcing amplitudes (i.e. $\lambda_{\bm{\beta},n}  = \lambda_{n} \sigma_n^{-2}$ for $n=1,2$). If the customary assumption of uniform forcing is applied, $\bm{\Lambda_{\bm{\beta}}} = \alpha \bm{I}$, then $\alpha  = \lambda_{1} \sigma_1^{-2} = \lambda_{2} \sigma_2^{-2}$ or, alternatively, $ \lambda_{1} / \lambda_{2} = \sigma_1^{2} /  \sigma_2^{2} $, and figure \ref{fig:Spectrums} shows this cannot be true. Therefore, unless the SPOD and resolvent spectra are equivalent, we must model or estimate the non-trivial forcing amplitudes.

The sizeable difference between the singular values reflects the forcing of different mechanisms at significantly different amplitudes in the flow. \cite{pickering2020lift} showed that there are three distinct spatial regions that lead to the most efficient amplification of the KH, Orr, and the lift-up mechanisms. They found that regions localized near the nozzle where perturbations are smaller, associate with KH-type responses, while regions downstream and near the end of the potential core where perturbations are significantly larger, support Orr-type responses. Considering these observations, a logical next step in completing a resolvent-based turbulence model is to tie the forcing amplitude of different modes to the turbulence intensities in the respective regions that force them.

\section{Conclusions}

We developed a data-informed optimization that quantitatively tested the extent to which an eddy-viscosity model improves agreement between observed large-scale structures, educed via SPOD, and those computed from resolvent analysis.
This eddy-viscosity approach acts as a proxy for modeling the effect of turbulence on large-scale structures and we found this approach provides substantial improvements in agreement (i.e. when compared to a baseline case that used a constant eddy-viscosity model corresponding to a value of $Re_T=3 \times 10^4$).  By directly optimizing the eddy viscosity field to achieve the best alignment, we found alignments between resolvent and SPOD modes as high as 96\%  or improvements of over 10-fold from the baseline alignment (i.e. 8\% to 80\%).

Across the frequencies and wavenumbers considered, the addition of an eddy-viscosity model to the resolvent operator highlighted its effect on the different amplifications mechanisms in the turbulent jet, Orr-type, KH-type, and lift-up. Although eddy-viscosity models improved modes related to the KH-type mode, we found KH modes to be rather insensitive to the eddy-viscosity field, a result expected from the inviscid nature of the inflectional KH instability.  For resolvent modes associated with the Orr and lift-up mechanisms, known to arise from nonlinear interactions, we found significant sensitivity. Resolvent modes computed without a sufficient eddy-viscosity model were visually unrecognizable from their SPOD counterpart, while those computed with an eddy-viscosity model aligned to nearly 80\%.

The optimal eddy-viscosity field also provided an upper bound for mode agreement, providing a benchmark to assess three additional eddy-viscosity models. Of these models, we found that traditional eddy-viscosity models (e.g. RANS based) perform nearly as well as the optimal eddy-viscosity models in aligning the most energetic mode. The traditional models even outperformed the optimal model (i.e. optimal in the first mode) when considering the subdominant modes, giving the greatest diagonalization of the forcing CSD at $m=0$, $St=0.6$ (i.e. ability to model the effect of nonlinear forcing).

Finally, we tested the modeling potential of a RANS-inferred EVRA through a sensitivity analysis and observed its performance over frequency, azimuthal wavenumber, and Mach number. We found the sensitivity of the RANS-based EVRA model calibration constant, $c$, to be weak, giving similar agreement for coefficients ranging over an order of magnitude. Choosing a frequency-independent ($c = 0.2$) RANS-EVRA model, we tested its performance across six azimuthal frequencies and three turbulent jets, spanning subsonic, transonic, and supersonic regimes. For the first three azimuthal wavenumbers (i.e. $m=0-2$), we observed substantially increased alignments for all three turbulent jets and across Strouhal numbers $St \in[0.05,1]$. Overall, these results show that ``classical’’ eddy-viscosity models (RANS or a mean-flow consistent model) aid in estimating the impact of the Reynolds stresses for resolvent analysis.

While the present data-driven analysis points to the efficacy of relatively simple eddy-viscosity-based models for modeling the effect of nonlinear forcing, there remains a need for refinements to this approach and careful comparison and consideration of alternative formulations. \ethan{Areas of interest include an investigation on the interpretability of the EVRA forcing structures and whether including an eddy-viscosity model hinders, or perhaps enhances, the use of the resolvent operator as a transfer function between known nonlinear forcings and their linear responses.}

\section*{Acknowledgments}
This research was supported by a grant from the Office of Naval Research (grant No. N00014-16-1-2445) with Dr. Steven Martens as program manager. E.P. was supported by the Department of Defense (DoD) through the National Defense Science \& Engineering Graduate Fellowship (NDSEG) Program. The LES study was performed at Cascade Technologies, with support from ONR and NAVAIR SBIR project, under the supervision of Dr. John T. Spyropoulos. The main LES calculations were carried out on DoD HPC systems in ERDC DSRC.

\section*{Declaration of interests}
The authors report no conflict of interest.

\appendix

\section{Linear damping term} \label{App:Linear}

Besides the studied eddy-viscosity models, we also investigated the impact of a linear damping term, which is equivalent to a finite-time-horizon resolvent analysis introduced by \cite{jovanovic2004modeling}, recently studied by \cite{yeh2019resolvent} to localize the resolvent forcing and response modes on an airfoil. For this model, we modify the operator so that,
\begin{align}
    \bm{L}_\beta = \bm{L} - \beta \bm{I},
\end{align}
where $\beta = 1/\tau > 0$, and $\tau$ is the desired temporal decay rate. We then find the value of $\beta$ that best aligns the dominant resolvent and SPOD modes.

\begin{figure}
	\centering
	\vspace{0.5cm}
	\includegraphics[width=0.5\textwidth]{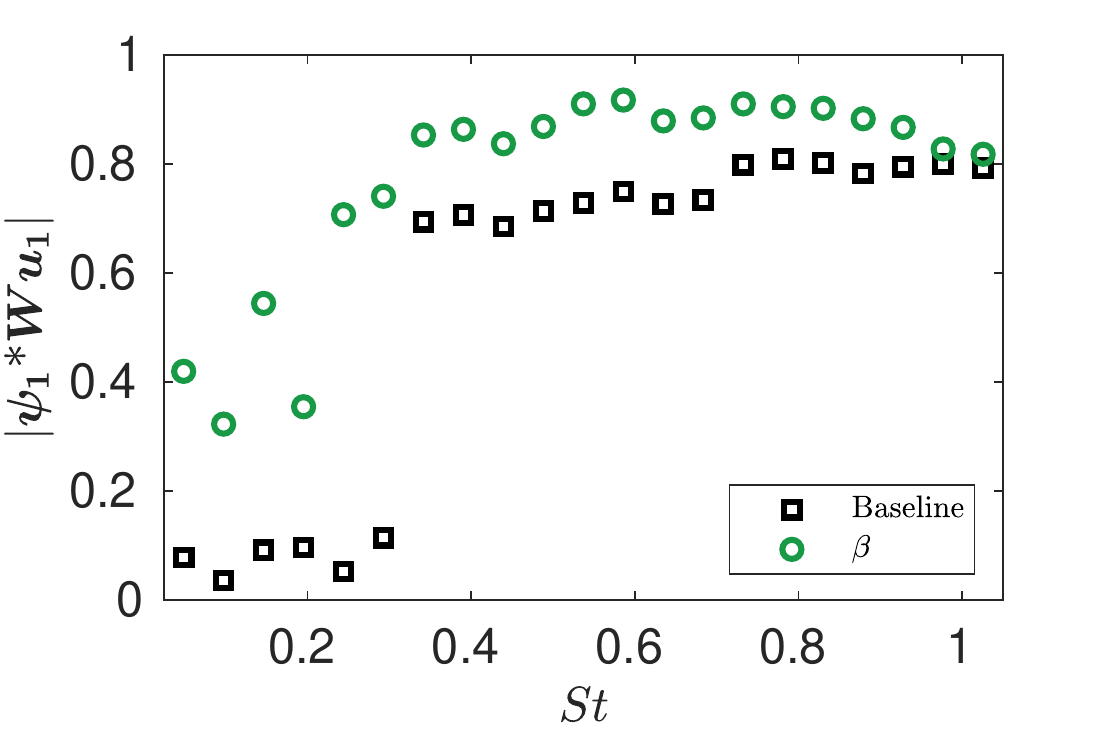}
	\caption{Optimal alignments for the linear damping term and the baseline case, $Re_T = 3 \times 10^4$.}.
	\label{fig:beta_alignments}
\end{figure}

Figure \ref{fig:beta_alignments} presents the alignments for the linear damping case. Although linear damping improves alignments, the performance is significantly inferior to the eddy-viscosity models, likely because of its monolithic damping effect over all wavenumbers, whereas the eddy-viscosity methods directly address the effect of the Reynolds stresses. Considering its suboptimal performance when compared to eddy-viscosity models, we only present results for the $M_j = 0.4$, $m=0$, and $St  \in[0.05,1]$ cases.

\section{Governing equations} \label{App:LNS}

Conservation of mass, momentum, and energy for a compressible, Newtonian fluid are written as,
\begin{align}
    \frac{D \rho}{D t} & = -\rho \Theta  \\
    \rho \frac{D \bm{u}}{D t} & =-\frac{1}{\gamma M_j^2}  \nabla(\rho T)+ \nabla \cdot \bigg[ \mu \bigg((\nabla \bm{u}) +(\nabla \bm{u})^T - {2 \over 3} \Theta \mathbb{I} \bigg) \bigg]\\
    \rho \frac{D T}{D t} & = -\frac{1}{\gamma M_j^2} \rho T \Theta + \frac{\mu}{(\gamma -1)  M_j^2 Pr_{\infty}} \nabla^2 T  \nonumber \\
    & + \gamma M_j^2 \mu \bigg[ \frac{1}{2} \bigg\{ (\nabla \bm{u}) + (\nabla \bm{u})^T \bigg\}:\bigg\{ (\nabla \bm{u}) + (\nabla \bm{u})^T \bigg\} - {2 \over 3} \Theta^2 \bigg],
\end{align}
respectively, where $\Theta = \nabla \cdot \bm{u}$ is the dilatation.  We take  $Pr_{\infty} = 0.7$ and $\gamma = 1.4$ as constants.  The equations have been made nondimensional with the jet density ($\rho_j$), speed ($U_j$) , and diameter, $D$.  The nondimensional viscosity, $\mu = {1 \over Re_j}$, is also a constant. 

Applying the Reynolds decomposition (i.e. $\bm{q} (\bm{x}, t) = \bar{\bm{q}}(\bm{x}) + \bm{q}^\prime(\bm{x}, t)$) to the above equations and separating terms that are linear and nonlinear in the fluctuations to the left- and right-hand sides, respectively, gives
\begin{align}
\frac{\bar{D} \rho^{\prime}}{D t} + \bm{u}^{\prime} \cdot \nabla \bar{\rho}+\rho^{\prime}\bar{\Theta}+\bar{\rho} \Theta^{\prime} &= f_{\rho} \label{eq:app_lin_rho} \\ 
\bar{\rho} \frac{ \bar{D} \bm{u}^{\prime}}{D t} + \bar{\rho} \bm{u}^{\prime} \cdot \nabla \overline{\bm{u}} 
+\rho^{\prime} \overline{\bm{u}} \cdot \nabla \overline{\bm{u}} \ & \nonumber \\ + \frac{1}{\gamma M_j^2}\left(\bar{\rho} \nabla T^{\prime}+\rho^{\prime} \nabla \bar{T}+\bar{T} \nabla \rho^{\prime}+T^{\prime} \nabla \bar{\rho}\right) & \nonumber \\ -
\nabla \cdot \bigg[ \mu \bigg((\nabla \bm{u}^{\prime})  +(\nabla \bm{u}^{\prime})^T - {2 \over 3} \Theta^{\prime} \mathbb{I} \bigg) \bigg]  &= \bm{f}_{\bm{u}} \\
\bar{\rho} \frac{\bar{D} T^{\prime}}{D  t}\bar{\rho} \mathbf{u}^{\prime} \cdot \nabla \bar{T}+\frac{1}{\gamma M_j^2}\left(\bar{T} \Theta^{\prime}+T^{\prime} \bar{\Theta}\right) & \nonumber \\ + \rho^{\prime}\{\overline{\mathbf{u}} \cdot \nabla \bar{T}+\frac{\bar{\rho}}{\gamma M_j^2} \bar{T} \bar{\Theta}\} \nonumber -\frac{\mu}{(\gamma -1)  M_j^2 Pr_{\infty}} \nabla^{2} T^{\prime} & \nonumber\\
    - \gamma M_j^2 \mu \left[\left\{(\nabla \overline{\mathbf{u}})+(\nabla \overline{\mathbf{u}})^{\mathrm{T}}\right\}:\left\{\left(\nabla \mathbf{u}^{\prime}\right)+\left(\nabla \mathbf{u}^{\prime}\right)^{\mathrm{T}}\right\}-\frac{4}{3} \bar{\Theta} \Theta^{\prime}\right] & = f_T. \label{eq:app_lin_T} 
\end{align}
with $\frac{\bar{D}}{D t} = {\partial \over \partial t} + \overline{\bm u} \cdot \nabla$, and where we have grouped all the nonlinear terms as forcing terms on the right-hand sides.

The left-hand-side is then transformed to a cylindrical coordinate frame and Fourier transformed in time $(\omega)$ and azimuth ($m$).  The resulting equations are discretized as discussed in \S~\ref{sec:methods}. 

The eddy-viscosity model we use, discussed in \S~\ref{sec:methods}, simply replaces $\mu$ in equations~(\ref{eq:app_lin_rho}) to (\ref{eq:app_lin_T}) with $\mu + \mu_T(x,r)$.  

\Revthree{
\section{Optimizing in an input and output framework} \label{App:inputoutput}} 

In resolvent analysis, it is often useful to restrict the input and output spaces by writing
\begin{align}
    \bm{L}_T \bm{q} &= \bm{B}\bm{f}\\ \nonumber
    \bm{y} &= \bm{C}\bm{q} \\
    \bm{y} &= \bm{C} \bm{L}_T^{-1} \bm{B} \bm{f}
\end{align}
where $\bm{C}$ transforms the state vector to a desired output space $\bm{y}$ and $\bm{B}$ maps a smaller dimensional input space to the state space. Here we show that such additions do not hinder the generality of the optimization presented in this manuscript.

The structure of the cost function does not change,
 \begin{equation}
         \mathcal{J} = \bm{u}_1^* \bm{W_y} \bm{\psi}_1 \bm{\psi}_1^* \bm{W_y} \bm{u}_1 - l^2\bm{\mu}_T^*\bm{M}{\bm{\mu}_T} + c.c,
 \end{equation}
 but the SPOD modes, $\bm{\psi}$, and the resolvent modes, $\bm{u}$, are now computed considering the observable $\bm{y}$ and the appropriate norms for the input and output space are defined by including weighting matrices $\bm{W_y}$ and$\bm{W}_f$, respectively.  The Lagrangian functional also takes a similar form as \S~\ref{sec:Opt_eddy}, 
\begin{align}
    \mathcal{L} & =  \bm{u}_1^* \bm{W_y} \bm{\psi}_1 \bm{\psi}_1^* \bm{W_y} \bm{u}_1-l^2\bm{\mu}_T^*\bm{M}{\bm{\mu}_T} -   \tilde{\bm{u}}_1^* \big(\bm{u}_1 - \bm{C} \bm{L}_T^{-1} \bm{B} \bm{v}_1\big) \\ \nonumber
    & -  \tilde{\bm{v}}_1^* \big(\bm{B}^* \bm{L}_T^{-*} \bm{C}^* \bm{W_y} \bm{u}_1 - \sigma_1^2 \bm{W_f} \bm{v}_1 \big) - \tilde{\sigma}_1 \big(\bm{u}_1^* \bm{W_y} \bm{u}_1 - 1\big) + c.c,
\end{align}
where, $\tilde{\bm{u}}_1, \tilde{\bm{v}}_1, \tilde{\bm{\sigma}}_1$ are the Lagrange multipliers. The effective composition of the functional is identical to that of the full-state optimization as it is composed of the cost function, the forward solution, the resolvent eigenvalue problem, and a normalization constraint. 
Taking variations with respect to each variable, with exception to the eddy-viscosity term, results in the following system of equations, 
\begin{align}
      \begin{bmatrix}
     \bm{I} & \bm{W_y}^*\bm{C}\bm{L}_T^{-1}\bm{B}  &  \bm{W_y} \bm{u}_1 \\
     \bm{B}^*\bm{L}_T^{-*} \bm{C}^*  &  \sigma_1^2 \bm{W_f}^* & 0 \\
     0 & \bm{v}_1^* \bm{W_f} & 0
     \end{bmatrix}
     \begin{bmatrix} \tilde{\bm{u}}_1\\ \tilde{\bm{v}}_1 \\ \tilde{\sigma}_1
     \end{bmatrix} 
     & =       
     \begin{bmatrix}  2 \bm{W_y} \bm{\Psi}_1 \bm{W_y} \bm{u}_1 \\ 0 \\ 0 \end{bmatrix}, \label{eqn:inputoutputsystem}
\end{align}
whose solution provides the Lagrange multipliers, $\tilde{\bm{u}}_1, \tilde{\bm{v}}_1, \tilde{\sigma}_1$. 

A difficulty that arises in building equation \ref{eqn:inputoutputsystem} is that the term $\bm{L}^{-1}_T$ is a large, dense matrix. When $\bm{L}$, $\bm{B}$, $\bm{C}$, and the weighting matrices are sparse, we may instead introduce auxiliary variables through
$$\bm{L}_T \tilde{\bm{\eta}}_1 = \bm{B} \tilde{\bm{u}}_1$$
$$\bm{L}_T^{*} \tilde{\bm{\zeta}}_1 = \bm{C}^* \tilde{\bm{v}}_1$$,
whereupon equation \ref{eqn:inputoutputsystem} may be written as a larger, but now sparse, system of equations
\begin{align}
\begin{bmatrix}
\bm{L}_T^* &  0 & -\bm{C}^* & 0 & 0\\ 
0 & \bm{L}_T &  0 & -\bm{B}  &  0\\ 
0 & \bm{W_y}^*\bm{C} & \bm{I} & 0 & \bm{W_y} \bm{u}_1 \\ 
\bm{B}^* &  0 & 0  & \sigma_1^2 \bm{W_f}^*  & 0 \\ 
0 & 0 & 0 & \bm{v}_1^* \bm{W_f} & 0
\end{bmatrix}\begin{bmatrix}
\tilde{\bm{\zeta}}_1 \\ 
\tilde{\bm{\eta}}_1 \\ 
\tilde{\bm{u}}_1\\ 
\tilde{\bm{v}}_1\\ 
\tilde{\sigma}_1
\end{bmatrix}=\begin{bmatrix}
0\\ 
0\\ 
 2 \bm{W_y} \bm{\Psi}_1 \bm{W_y} \bm{u}_1\\ 
0\\ 
0
\end{bmatrix}.
\end{align}

The above presents a general optimization framework for aligning any input-output resolvent analysis to data (i.e. here we use SPOD modes, but $\bm{\Psi}$ need not be restricted to SPOD modes). Variations with respect to any parameter of the resolvent operator may now be made to investigate their effect on modelling (or assimilating) known quantities. 

\bibliographystyle{jfm}
\bibliography{jfmbib}

\end{document}